\documentclass[superscriptaddress,aps,pra,twocolumn,showkeys]{revtex4-2}
\usepackage{amsmath,mathtools,siunitx}
\usepackage{amssymb}
\usepackage{graphicx}
\usepackage{color}
\usepackage{bm}
\usepackage[colorlinks,linkcolor=blue,anchorcolor=blue,citecolor=blue,urlcolor=blue]{hyperref}
\usepackage{booktabs}
\usepackage{braket} 
\usepackage{bm}
\usepackage{tabularx}
\usepackage{titlesec} 
\usepackage{float}

\begin{document}
\title{Decoherence-free interaction and maximally entangled state generation in giant-atom semi-infinite waveguide systems}

\author{Jie Liu}
\affiliation{Lanzhou Center for Theoretical Physics, Key Laboratory of Theoretical Physics of Gansu Province, Key Laboratory of Quantum Theory and Applications of MoE, Gansu Provincial Research Center for Basic Disciplines of Quantum Physics, Lanzhou University, Lanzhou 730000, China}

\author{Yue Cai}
\affiliation{Lanzhou Center for Theoretical Physics, Key Laboratory of Theoretical Physics of Gansu Province, Key Laboratory of Quantum Theory and Applications of MoE, Gansu Provincial Research Center for Basic Disciplines of Quantum Physics, Lanzhou University, Lanzhou 730000, China}

\author{Lei Tan}
\email{tanlei@lzu.edu.cn}
\affiliation{Lanzhou Center for Theoretical Physics, Key Laboratory of Theoretical Physics of Gansu Province, Key Laboratory of Quantum Theory and Applications of MoE, Gansu Provincial Research Center for Basic Disciplines of Quantum Physics, Lanzhou University, Lanzhou 730000, China}

\begin{abstract}
Giant atoms are artificial atoms that can couple to a waveguide non-locally. Previous works have shown that two giant atoms in a braided configuration can interact through one-dimensional (1D) infinite and chiral waveguides, with both individual and collective atomic relaxation being fully suppressed. In this paper, however, we show that the decoherence-free interaction (DFI) between two giant atoms can be realized in both braided and nested configurations when the waveguide is semi-infinite. This protected interaction fails to appear in semi-infinite waveguide systems containing two separate giant atoms or two small atoms.
We also study the entanglement generation between two giant atoms coupled to a 1D semi-infinite waveguide. The results show that the maximally entangled state is generated in both braided and nested configurations due to the formation of DFI, and in the separate configuration, the maximally achievable entanglement can exceed 0.5. Finally, we generalize the discussion on DFI and entanglement generation to the case involving multiple giant atoms coupled
into a semi-infinite waveguide.
This study presents a new scheme for realizing DFI and generating maximally entangled states in giant-atom waveguide-QED systems.

\end{abstract}

\maketitle
\section{INTRODUCTION}

Waveguide quantum electrodynamics (QED) systems are an important platform for studying atom-photon interactions and have attracted much attention in recent years \cite{Gu7182017}. Many physical phenomena, such as single-photon transport \cite{Shen95213001, Shen79023837, Liao80014301, Liao92023806}, two-photon transport \cite{Hu97033847, Liao82053836}, entanglement generation between distant atoms \cite{Gonzalez24, Cano25, Carlos26, Gonzalez27, Paolo28, Gonzalez92}, and superradiant and subradiant state creation \cite{DeVoe76, Zanner2022, vanLoo2013}, have been observed in waveguide-QED systems. In these reports, atoms are typically idealized as point-like objects and thus they only interact with the waveguide at single points.

Recently, waveguide-QED systems containing giant atoms have garnered significant attention from many peers. In such systems, giant atoms are artificial atoms that can interact with the waveguide at multiple connection points \cite{Kockum2021}. Several interesting quantum optical phenomena have been demonstrated in giant-atom waveguide-QED systems, including frequency-dependent relaxation rates and Lamb shifts \cite{Kockum46}, non-Markovian decay dynamics \cite{Guo95, Du103, Du4}, single-photon scattering \cite{Zhao101, Cai104, Yin106}, and DFI \cite{Kockum47, Cilluffo49, Carollo50, Soro105, Kannan48, Soro107013710, Du107023705, Raaholt6043222}. In addition, the generation \cite{Yin55, Yin56, 130, Luo57, liu2024} and transfer \cite{Liu2025} of quantum entanglement have also been investigated in giant-atom waveguide-QED systems. 

However, it is important to note that in most studies on giant atoms, the waveguide is usually considered infinite or chiral. A key question is how giant atoms behave when coupled to a semi-infinite waveguide. The semi-infinite waveguide is a waveguide with one end terminated by a perfect mirror \cite{Zhang101032335, Xin105053706, Tufarelli90012113, Bradford87063830, Chang024305, Barkemeyer023708, Dorner023816, DingM1590482, Zeng30305, PKisa033704, Rubies213603, Eschner2001, Koshino2012}. Compared to the infinite and chiral waveguides, the semi-infinite waveguide is particularly notable in that the mirror at the end allows the atom to interact twice with the field in the waveguide. Hence, when the giant atom couples to the semi-infinite waveguide, its behavior may differ from that in the infinite and chiral waveguides.
While the semi-infinite waveguide setup containing giant atoms has been addressed in previous studies \cite{Kockum46, Li109, sun2025non}, only a single giant atom was considered. 

In this work, our study goes beyond the framework of a single giant atom coupled to a semi-infinite waveguide. We consider an arbitrary number of giant atoms $M$ interacting with a 1D semi-infinite waveguide. Treating the semi-infinite waveguide as the environment of atoms, we derive the Markov quantum master equation governing the dynamical evolution of $M$ giant atoms. 

We focus on the case of two giant atoms. Concretely, we analyze the behavior of two giant atoms coupled to a 1D semi-infinite waveguide with three configurations: braided, separate, and nested.
In doing so, we show that the DFI between two giant atoms, previously only observed within the braided configuration in 1D infinite \cite{Kockum47, Kannan48, Soro107013710, Du107023705} and chiral \cite{Cilluffo49, Carollo50, Soro105} waveguide systems, can be realized in both braided and nested configurations in 1D semi-infinite waveguide systems. This protected interaction is not supported in a semi-infinite waveguide setup involving either two separate giant atoms or two small atoms. For two small atoms coupled to a semi-infinite waveguide, it has been reported that there is no DFI between them \cite{Kockum47}. 

Furthermore, we study the entanglement generation between two giant atoms coupled to a semi-infinite waveguide with three configurations. We show that the maximally achievable entanglement for all three configurations can exceed 0.5. In particular, two giant atoms in braided and nested configurations can be in a maximally entangled state, which is attributed to the formation of DFI. While the maximally entangled state has been realized in the braided configuration in infinite waveguide systems \cite{Yin55}, 
\begin{figure}[!htbp]
	\includegraphics[width=0.48\textwidth]{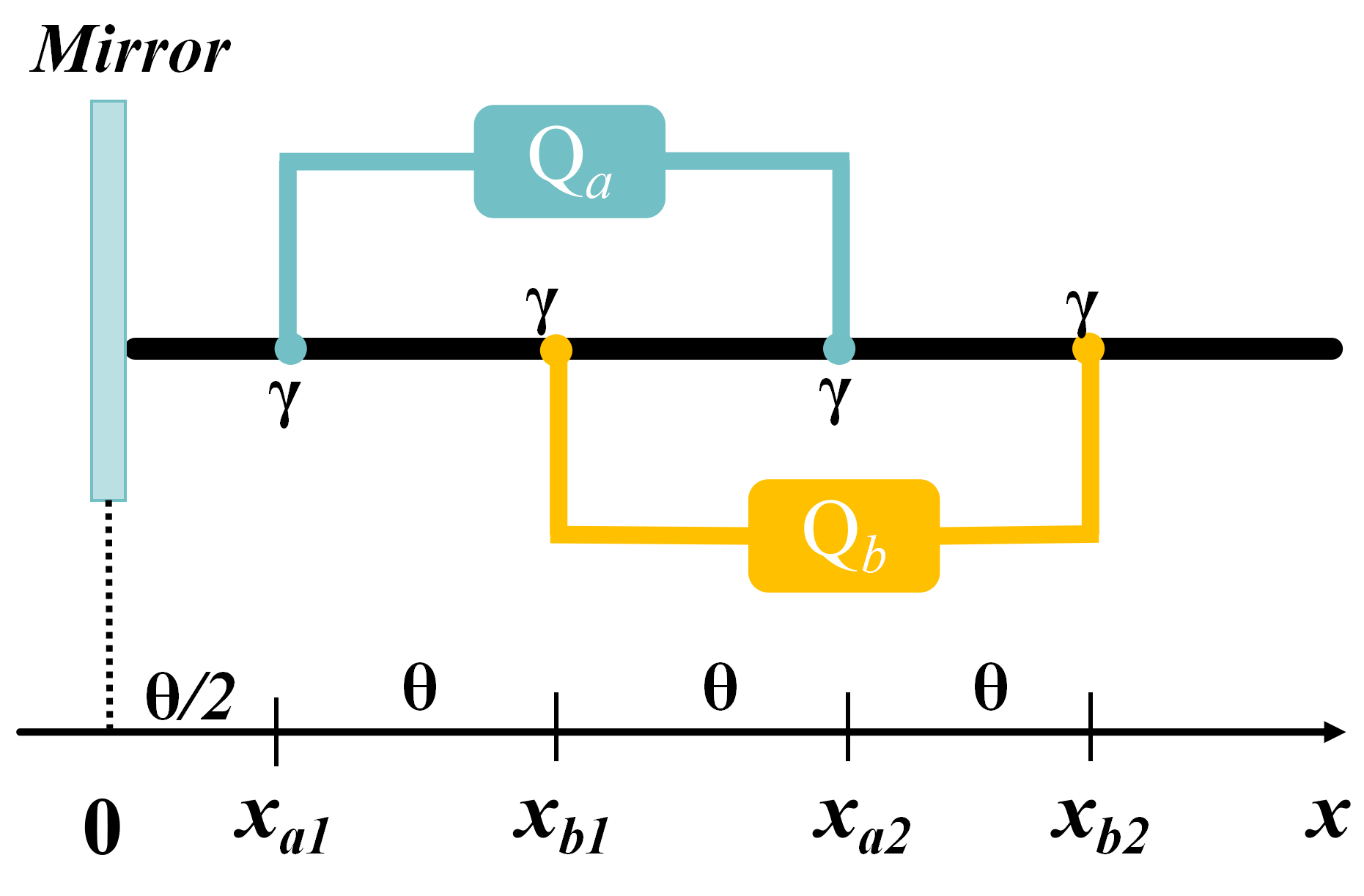}\hfill
	\caption{(Color online)  Two giant atoms in a braided configuration coupled to a 1D semi-infinite waveguide. The connection points are located at coordinates $x_{j1}$ and $x_{j2}$, where $j = a, b$. In our work, we consider the case where the two giant atoms share the same transition frequency, i.e., $\omega_a = \omega_b = \omega_0$.}
	\label{fig1}
\end{figure}
the ability to generate a maximally entangled state in the nested configuration and to generate an entangled state with entanglement exceeding 0.5 in the 
separate configuration is unique to semi-infinite waveguide systems.

Finally, we generalize the discussion on DFI and entanglement generation to the case where multiple giant atoms are coupled to a semi-infinite waveguide. The results demonstrate that, for multiple giant atoms in both braided and nested configurations, DFI can still be realized, and a high degree of entanglement between the atoms can be generated.

We believe that the DFI between giant atoms can find applications in various fields, e.g., in quantum simulation \cite{Georgescu153, bulutaquantum, Wang012334, trabesinger2012quantum, lloyd1996universal} and quantum metrology \cite{Giovannetti010401, giovannetti2011advances, xiang2013quantum}. 
In these fields, protecting qubits from decoherence while enabling their interaction is crucial. 
Furthermore, DFI has potential applications in quantum computation \cite{Galindo347, bennett2000quantum, nielsen2010quantum, divincenzo1995quantum, nature2022quantum}. For instance, future large-scale quantum computers could be operated within the DFI framework to protect the fragile quantum information during computation. It is also possible to use DFI to generate the maximally entangled state (as we discussed in this work), which is a core resource in quantum communication \cite{gisin2007quantum, cariolaro2015quantum} and quantum networks \cite{kimble2008quantum, wehner2018quantum}.

This paper begins with Sec. \ref{II}, which describes the model for giant atoms coupled to a 1D semi-infinite waveguide. Then, we specifically explore the DFI and entanglement generation between two braided giant atoms in Sec. \ref{III} and \ref{IV}. In Sec. \ref{V}, we consider the case of two giant atoms in the separate and nested configurations. In Sec. \ref{VI}, we discuss how the
results found here generalize to multiple atoms.
Finally, we discuss the entanglement generation between two giant atoms in the two-excitation subspace and the experimental feasibility of the proposed scheme in Sec. \ref{VII}, and summarize the main findings of this study in Sec. \ref{VIII}. In addition, we provide four appendices. In Appendix \ref{Appendix A}, the detailed derivation of the master equation is shown. In Appendix \ref{Appendix B}, we present the analytical expressions for the concurrence that depend on the phase shift. In Appendix \ref{Appendix C}, we replot the corresponding results for the infinite waveguide case, and relevant coefficients associated with separate and nested configurations in the semi-infinite waveguide system are provided in Appendix \ref{Appendix D}.

\section{THEORETICAL FRAMEWORK}\label{II}
We start by formulating a general model in which $M$ two-level giant atoms are coupled to a 1D semi-infinite waveguide along the positive $x$ semi-axis, with each atom $j$ having $N_j$ connection points. In this model, we assume the semi-infinite waveguide is realized by terminating the left end (coordinate set to $x=0$) of the 1D infinite waveguide with a mirror, and all atoms are located on the right side of the mirror.

Assuming that the coupling between giant atoms and the semi-infinite waveguide is weak at each connection point, and that the relaxation times of giant atoms are much longer than all other relevant time scales in the system, we derive the quantum master equation describing the dynamical evolution of $M$ giant atoms (see Appendix \ref{Appendix A} for detailed derivations). In the interaction picture, the master equation takes the following form (hereafter, \(\hbar = 1\)):
\begin{widetext}
	\begin{equation}
		\begin{aligned}
			\dot{\rho} = & -i \left[ \sum_{j=1}^{M} \delta \omega_{j} \sigma_{+}^{j}\sigma_{-}^{j} + \sum_{j=1}^{M-1} \sum_{k=j+1}^{M}  g_{j,k} (  \sigma_+^k \sigma_-^j + \text{H.c.} ), \rho \right] +
			\sum_{j=1}^{M} \Gamma_{j} \mathcal{D}[\sigma_-^j] \rho \\
			& + \sum_{j=1}^{M-1} \sum_{k=j+1}^{M} \Gamma_{coll,j,k} \left[  ( \sigma_-^j \rho \sigma_+^k - \frac{1}{2} \{ \sigma_+^k \sigma_-^j, \rho \})  + \text{H.c.} \right],
		\end{aligned}
		\label{eq:me1}
	\end{equation}
\end{widetext}
with
\begin{widetext}
	\begin{equation}
		\begin{aligned}
			\delta \omega_{j} &= \sum_{n=1}^{N_j} \sum_{m=1}^{N_j} \frac{\sqrt{\gamma_{jn} \gamma_{jm}}}{2}  \Big\{ \sin \left[k_0|x_{jn}-x_{jm}|\right] +  \sin \left[k_0(x_{jn}+x_{jm})\right] \Big\}, \\
			g_{j,k} &= \sum_{n=1}^{N_j} \sum_{m=1}^{N_k} \frac{\sqrt{\gamma_{jn} \gamma_{km}}}{2}  \Big\{\sin \left[k_0|x_{jn}-x_{km}|\right] + \sin \left[k_0(x_{jn}+x_{km})\right] \Big\}, \\
			\Gamma_{j} &= \sum_{n=1}^{N_j} \sum_{m=1}^{N_j} \sqrt{\gamma_{jn} \gamma_{jm}}  \Big\{ \cos \left[k_0|x_{jn}-x_{jm}|\right] +  \cos \left[k_0(x_{jn}+x_{jm})\right] \Big\}, \\
			\Gamma_{coll,j,k} &= \sum_{n=1}^{N_j} \sum_{m=1}^{N_k} \sqrt{\gamma_{jn} \gamma_{km}}  \Big\{\cos \left[k_0|x_{jn}-x_{km}|\right] + \cos \left[k_0(x_{jn}+x_{km})\right] \Big\},
		\end{aligned}
		\label{eq:ex1}
	\end{equation}
\end{widetext}
where $\delta \omega_{j}$ represents the Lamb shift of giant atom $j$, $\mathcal{D}[X]\rho = X \rho X^\dagger - \frac{1}{2} \{X^\dagger X, \rho\}$, $g_{{j,k}}$ is the exchange interaction between giant atoms, $\Gamma_{j}$ represents the individual decay rate of giant atom $j$, $\Gamma_{coll,j,k}$ is the collective decay rate for giant atoms, and H.c. denotes the Hermitian conjugate.
Moreover, $\gamma_{jn}$ represents the decay rate of atom $j$ at the connection point $n$ (coordinate $x_{jn}$), $k_0|x_{jn}-x_{jm}|$ corresponds to the phase acquired for a photon with wave vector $k_0$ traveling between connection points $n$ and $m$ of the giant atom $j$, and $k_0(x_{jn}+x_{jm})$ represents the phase acquired during the process in which the photon released from connection point $n$ of atom $j$ is absorbed by the connection point $m$ of the same atom after being reflected through the mirror located at $x=0$, and the remaining terms can be interpreted in a similar manner. 

From Eq. (\ref{eq:ex1}), the primary difference with respect to the infinite \cite{Kockum47, Kannan48} and chiral \cite{Soro105} waveguide cases is that the expressions for $\delta \omega_{j}$, $g_{j,k}$, $\Gamma_{j}$, and $\Gamma_{coll,j,k}$ no longer depend just on the relative distances between the connection points: they are also determined by the distance from the connection points to the mirror in the semi-infinite waveguide case.  

We focus on the case of two giant atoms where each atom interacts with the waveguide at two connection points. For two giant atoms with two connection points each, three distinct topology configurations are possible \cite{Kockum47, Soro105}: braided, separate, and nested. The setup we initially consider is the braided configuration, as shown in Fig. \ref{fig1}. Later, we consider the separate and nested configurations in Sec. \ref{V}.
To simplify the discussion, the model in Fig. \ref{fig1} adopts the following assumptions:
$(i)$ the decay rate at each connection point is identical and denoted by $\gamma$, and $(ii)$ the phase shift between adjacent connection points is defined as $\theta$, where $\theta/2$ represents the phase shift accumulated as a photon travels from position $x_{a1}$ to the mirror ($x=0$).
When the system contains two giant atoms, Eq. (\ref{eq:me1}) reduces to
\begin{equation}
	\begin{aligned}
		\dot{\rho} = & -i \left[ \delta \omega_{a} \sigma_{+}^{a}\sigma_{-}^{a} + \delta \omega_{b} \sigma_{+}^{b}\sigma_{-}^{b} + g_{a,b} \left(\sigma_{+}^{b}\sigma_{-}^{a} + \text{H.c.}\right), \rho \right] \\
		&+\Gamma_{a} \mathcal{D}[\sigma_{-}^{a}] \rho +\Gamma_{b} \mathcal{D}[\sigma_{-}^{b}] \rho   \\
		& + \Gamma_{coll,a,b} \left[ (\sigma_{-}^{a}\rho\sigma_{+}^{b} - \frac{1}{2} \left\{ \sigma_{+}^{b}\sigma_{-}^{a}, \rho \right\})+ \text{H.c.}\right].
	\end{aligned}
	\label{eq:me2}
\end{equation}
 For the two braided giant atoms shown in Fig. \ref{fig1}, the expressions of the coefficients in Eq. (\ref{eq:me2}) are given in the form below: 
 \begin{equation}
 	\begin{aligned}
 		\delta \omega_{a} &= \gamma \sin (2\theta) + \gamma \sin (3\theta) + \frac{\gamma}{2} \sin \theta + \frac{\gamma}{2} \sin (5\theta), \\
 		\delta \omega_{b} &= \gamma \sin (2\theta) + \gamma \sin (5\theta) + \frac{\gamma}{2} \sin (3\theta) + \frac{\gamma}{2} \sin (7\theta), \\
 		g_{a,b} &= \frac{\gamma}{2} ( 3\sin \theta + \sin (3\theta) + \sin (2\theta) + 2 \sin (4\theta) \\
 		&\quad+ \sin (6\theta) ), \\
 		\Gamma_{a} &= 2\gamma + 2\gamma \cos(2\theta) + 2\gamma \cos(3\theta) + \gamma \cos \theta \\
 		&\quad+ \gamma \cos(5\theta), \\
 		\Gamma_{b} &= 2\gamma + 2\gamma \cos(2\theta) + 2\gamma \cos(5\theta)+\gamma\cos(3\theta)\\  &\quad + \gamma \cos(7\theta), \\
 		\Gamma_{coll,a,b} &= \gamma (3 \cos\theta + \cos(3\theta) + \cos(2\theta) + 2\cos(4\theta) \\
 		&\quad+ \cos(6\theta) ).
 	\end{aligned}
 	\label{eq:ex2}
 \end{equation}
The more general forms of the coefficients, which account for arbitrary decay rates at each connection point and arbitrary phase shifts between connection points, are provided in Appendix \ref{Appendix A}.

\begin{figure}[!htbp]
	\includegraphics[width=0.45\textwidth]{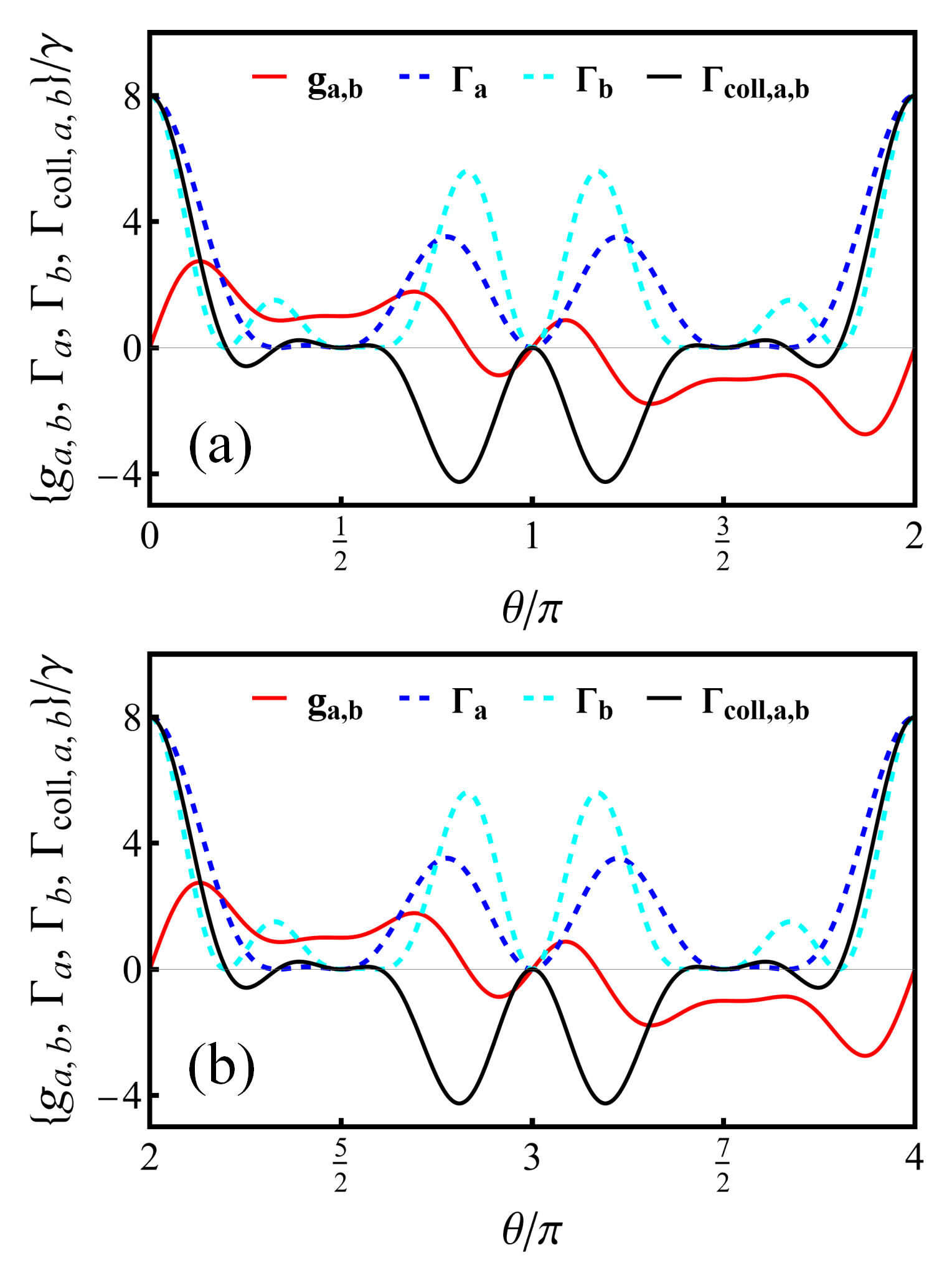}\hfill
	\caption{(Color online) Exchange interaction and decay rates as a function of $\theta/\pi$ for two braided giant atoms.}
	\label{fig2}
\end{figure}
\section{DFI}\label{III}
In previous works, it has been confirmed that two braided giant atoms can interact via 1D infinite \cite{Kockum47, Kannan48, Soro107013710, Du107023705} and chiral \cite{Cilluffo49, Carollo50, Soro105} waveguides without losing their excitations to the waveguide environment. That is, there is a DFI between two braided giant atoms. 
Mathematically, the condition for DFI to manifest can be stated as
\begin{equation}
	\begin{aligned}
		g_{a,b} &\neq 0, \\
		\Gamma_{a} = \Gamma_{b} &= \Gamma_{coll,a,b} = 0.
	\end{aligned}
	\label{eq:dfcondition}
\end{equation}

In this work, we show that the DFI between two braided giant atoms remains possible when the waveguide is semi-infinite. In Fig. \ref{fig2} we plot the exchange interaction $g_{{a,b}}$ and decay rates $\Gamma_{a}$, $\Gamma_{b}$, $\Gamma_{coll,a,b}$ from Eq. (\ref{eq:ex2}) as a function of $\theta/\pi$. Specifically, Fig. \ref{fig2}$(a)$ shows the results for $ \theta \in [0, 2\pi]$, while Fig. \ref{fig2}$(b)$ shows the results for $ \theta \in [2\pi, 4\pi]$.  
By comparing Fig. \ref{fig2}$(a)$ with Fig. \ref{fig2}$(b)$, we conclude that the exchange interaction $g_{{a,b}}$ and the decay rates $\Gamma_{a}$, $\Gamma_{b}$, $\Gamma_{coll,a,b}$ are phase dependent with a period of 2$\pi$. Therefore, we restrict our discussion to Fig. \ref{fig2}$(a)$, which covers one full period.

From Fig. \ref{fig2}$(a)$, we can observe that when $\theta = \pi/2$ and $3\pi/2$, the decay rates $\Gamma_{a}$, $\Gamma_{b}$, $\Gamma_{coll,a,b}$ are all \textit{zero} while the exchange interaction $g_{{a,b}}$ remains \textit{nonzero}.
This implies that DFI can be realized between two braided giant atoms in the semi-infinite waveguide system.

The above analysis on DFI is naturally extended to the general case of $ \theta \in [n\pi, (n+1)\pi]$ (where $n$ is an integer). Based on the periodic behavior of the system, we easily find that in the region of $ \theta \in [n\pi, (n+1)\pi]$, there exists DFI when
\begin{equation}
	\theta = \left(n + \frac{1}{2}\right)\pi.
	\label{eq:ex11}
\end{equation}

\section{ENTANGLEMENT GENERATION BETWEEN TWO BRAIDED GIANT ATOMS}\label{IV}
This section explores the entanglement generation between two braided giant atoms, where each giant atom is coupled to the semi-infinite waveguide at two distinct connection points. For two two-level atoms, the entanglement between them can be quantified by concurrence \cite{Wootters61}. In the following, we first derive the expression for the concurrence of the two-giant-atom system.
\subsection{Concurrence} \label{IV A}
In our work, we focus on entanglement generation in the case where only one giant atom is excited. When there is only one excitation in the system, we can remove the jump term (i.e., $\sigma_-^a \rho \sigma_+^a, \sigma_-^b \rho \sigma_+^b, \sigma_-^a \rho \sigma_+^b$, and $\sigma_-^b \rho \sigma_+^a$, which describe the effect of the measurement on
the state of the system \cite{Wiseman2010, Haroche2006, Minganti2019}) in the quantum master equation (\ref{eq:me2}) to obtain a non-Hermitian effective Hamiltonian $H_{\text{eff}}$ \cite{Yin55, liu2024, Mok101, JinHuiWu}:
\begin{equation}
	\begin{aligned}
		{H}_{\text{eff}} &= \sum_{j=a,b} \delta \omega_{j} {\sigma}_+^j {\sigma}_-^j 
		+ g_{{a,b}} ({\sigma}_+^b {\sigma}_-^a + {\sigma}_+^a {\sigma}_-^b) \\
		&\quad - \frac{i}{2} \sum_{j=a,b} \Gamma_{j} {\sigma}_+^j {\sigma}_-^j 
		- \frac{i}{2} \Gamma_{coll,a,b} ({\sigma}_+^b {\sigma}_-^a + {\sigma}_+^a {\sigma}_-^b).
	\end{aligned}
	\label{eq:Heff}
\end{equation}

In this situation, the Schrödinger equation with the non-Hermitian Hamiltonian $H_{\text{eff}}$ can effectively describe the dynamic evolution of the two-giant-atom system
\begin{equation}
	i \frac{\partial}{\partial t} |\psi(t)\rangle = {H}_{\text{eff}} |\psi(t)\rangle,
	\label{eq:Sequation}
\end{equation}
where $|\psi(t)\rangle$ is the state vector of the system:  
\begin{equation}
	|\psi(t)\rangle = c_{eg}(t) |e\rangle_a |g\rangle_b + c_{ge}(t) |g\rangle_a |e\rangle_b.
	\label{eq:Wf}
\end{equation}

Then, the density matrix for the two giant atoms in the basis $\left\{|e\rangle_a |e\rangle_b, |e\rangle_a |g\rangle_b, |g\rangle_a |e\rangle_b, |g\rangle_a |g\rangle_b \right\}$ can be obtained from $\rho(t)= |\psi(t)\rangle \langle \psi(t)|$ 
\begin{equation}
	\label{pmatrix}
	\begin{aligned}
		\rho(t) &= \begin{pmatrix}
			0 & 0 & 0 & 0 \\
			0 & |{c}_{eg}(t)|^2 & {c}_{eg}(t) {c}_{ge}^*(t) & 0 \\
			0 & {c}_{eg}^*(t) {c}_{ge}(t) & |{c}_{ge}(t)|^2 & 0 \\
			0 & 0 & 0 & 0
		\end{pmatrix}.
	\end{aligned}
\end{equation}

For this density matrix, the concurrence is given by \cite{Wootters61}:
\begin{equation}
	C(t) =2|c_{eg}(t) c_{ge}^{\ast }(t)|.\label{eq:concurrence}
\end{equation}
Note that when $C = 0$, the state is separable, and when $C = 1$, the state is maximally entangled. In this section, and in the part of Sec. \ref{V} that concerns entanglement generation, we let the two giant atoms start in the state $|e\rangle_a |g\rangle_b$.
According to this initial condition, expressions for probability amplitudes $c_{\text{eg}}(t)$ and $c_{\text{ge}}(t)$ can be obtained by solving Eq. (\ref{eq:Sequation}) and then the entanglement of two giant atoms can be measured through Eq. (\ref{eq:concurrence}).
\subsection{Entanglement generation}\label{IV B}
We now study the entanglement generation between two braided giant atoms coupled to a semi-infinite waveguide.
Based on Eqs. (\ref{eq:Heff})-(\ref{eq:concurrence}), the analytical form of concurrence is derived as follows:
\begin{equation}
	C_{eg}^{B}(t) = \frac{e^{-\frac{1}{2} \, \mathrm{Re}\left[ R_1(t) \right]}}{2} 
	\left| \frac{R_2(t)\, \left( R_3(t) + R_4(t) \right)}{A A^*} \right|.
	\label{eq:Concurrence}
\end{equation}
where the superscript $B$ and subscript $eg$ denotes the braided configuration case and atomic initial state $|e\rangle_a |g\rangle_b$, respectively. Similarly, superscripts $S$ and $N$, used in Sec. \ref{V}, correspond to the separate and nested configurations. The expressions for $A$ and $R_{i}(t)$ ($i=1,2,3,4$), which depend on the phase shift $\theta$, are given in Appendix \ref{Appendix B}.

In Fig. \ref{fig3}$(a)$, we show the evolution of concurrence $C_{eg}^{B}$ as a function of $\theta/\pi$ and $\gamma t$. From Fig. \ref{fig3}$(a)$, we can see that concurrence $C_{eg}^{B}$ is phase dependent with a period of 2$\pi$. This contrasts with the infinite waveguide case \cite{Yin55}, where the evolution of the concurrence with $\theta$ is characterized by a periodicity of $\pi$. To enable a direct comparison, the entanglement dynamics of two braided giant atoms in the infinite waveguide case, reported in Ref.~\cite{Yin55}, are replotted and presented in Appendix~\ref{Appendix C}. Moreover, Fig. \ref{fig3}$(a)$ shows that for $\theta \in [0, \pi]$, the relation $C_{eg}^{B}(t, \theta) = C_{eg}^{B}(t, 2\pi- \theta)$ holds. 

To clearly elucidate the phase-shift dependence of entanglement generation, we now analyze the evolution of the concurrence $C_{eg}^{B}$ when phase shift $\theta$ takes some specific values, where $\theta \in [0, \pi]$. 
As shown in Fig. \ref{fig3}$(a)$, when $\theta = 0$, the concurrence $C_{eg}^{B}$ exhibits steady-state behaviors. 
\begin{figure}[!htbp]
	\includegraphics[width=0.45\textwidth]{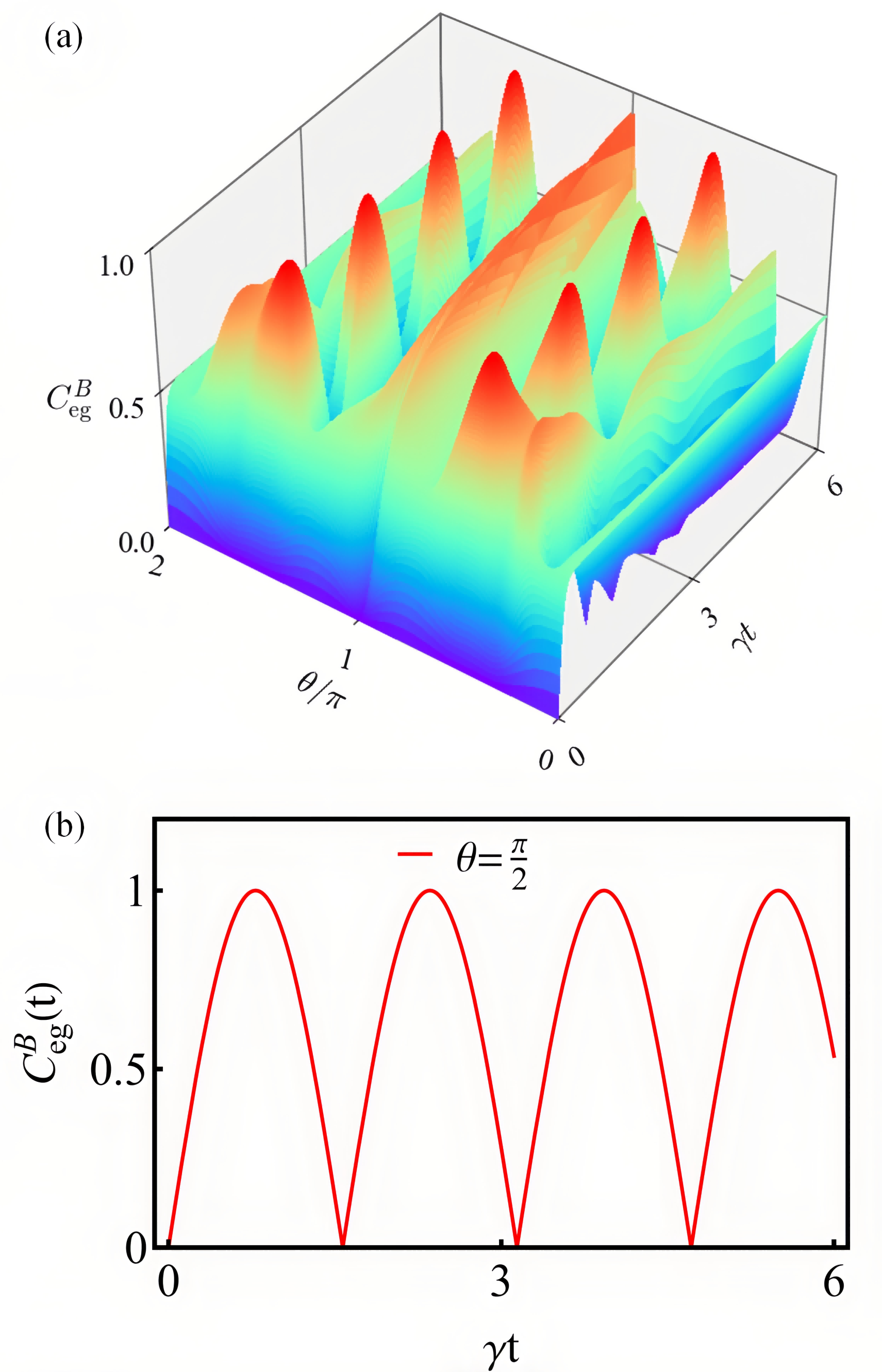}\hfill
	\caption{(Color online)  $(a)$ Concurrence $C_{eg}^{B}$ as a function of $\theta/\pi$ and $\gamma t$. $(b)$ Concurrence $C_{eg}^{B}(t)$ as a function of $\gamma t$ at a given value of phase shift. }
	\label{fig3}
\end{figure}
The steady-state behavior of the concurrence $C_{eg}^{B}$ at $\theta = 0$ can be explained through the 
analytical expression. When $\theta = 0$, the concurrence in Eq. (\ref{eq:Concurrence}) becomes
\begin{equation}
C_{eg}^{B}(t) = \frac {1}{2} \left({1 - e^{-16 \gamma t}}\right).
	\label{eq:ex6}
\end{equation}
From Eq. (\ref{eq:ex6}), we find that the concurrence $C_{eg}^{B}(t)$ converges to a steady state value 0.5 with a rate of $16\gamma$. We note that this convergence rate differs from that in the infinite waveguide case, where the rate is 8$\gamma$ \cite{Yin55}.
For the phase shift $\theta = \pi/2$, where DFI is formed, the concurrence $C_{eg}^{B}$ is found to periodically oscillate between zero and one, a behavior identical to that found in the infinite waveguide setup \cite{Yin55}, as presented in Appendix~\ref{Appendix C}. This oscillatory behavior is illustrated more clearly in Fig. \ref{fig3}$(b)$. Mathematically, the oscillatory behavior observed at $\theta = \pi/2$ can be explained by reducing Eq. (\ref{eq:Concurrence}) to the following form: 
\begin{equation}
	C_{eg}^{B}(t)=|\sin(2\gamma t)|.\label{Concurrence3}
\end{equation}
According to Eq. (\ref{Concurrence3}), it can be found that the concurrence $C_{eg}^{B}(t)$ oscillates between zero and one with a period $\frac{\pi}{2\gamma}$. When $t=\frac{n\pi}{4\gamma}$ (where $n$ is an odd integer), concurrence $C_{eg}^{B}(t)=1$, indicating that the two giant atoms reach a maximally entangled state.
Specifically, when $\theta$ is taken to be $\pi$, no entanglement is generated here, i.e., $C_{eg}^{B} = 0$. The reason is mainly that when $\theta = \pi$, exchange interaction $g_{a,b}$, individual decay rates $\Gamma_{a}$ and $\Gamma_{b}$, as well as the collective decay rate $\Gamma_{\text{coll},a,b}$ all become zero (as can be seen from Fig. \ref{fig2}), which implies that the two giant atoms decouple from the waveguide, thereby making entanglement generation impossible.
\begin{figure}[!htbp]
	\includegraphics[width=0.48\textwidth]{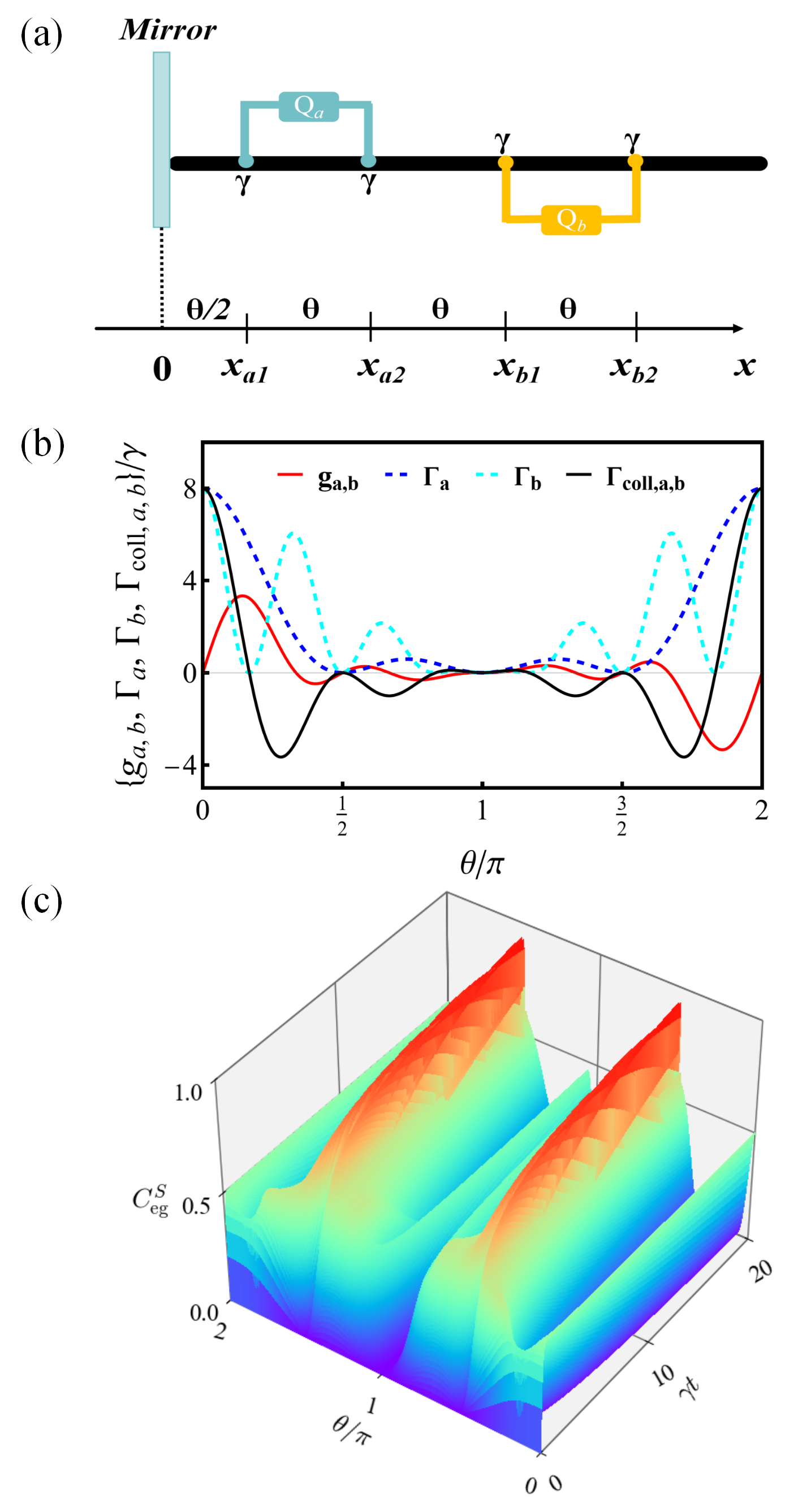}\hfill
	\caption{(Color online)  $(a)$ Two giant atoms in a separate configuration coupled to a 1D semi-infinite waveguide. $(b)$ Exchange interaction and decay rates as a function of $\theta/\pi$ for two separate giant atoms. $(c)$ Concurrence $C_{eg}^{S}$ as a function of $\theta/\pi$ and $\gamma t$. }
	\label{fig4}
\end{figure}

\begin{figure}[!htbp]
	\includegraphics[width=0.48\textwidth]{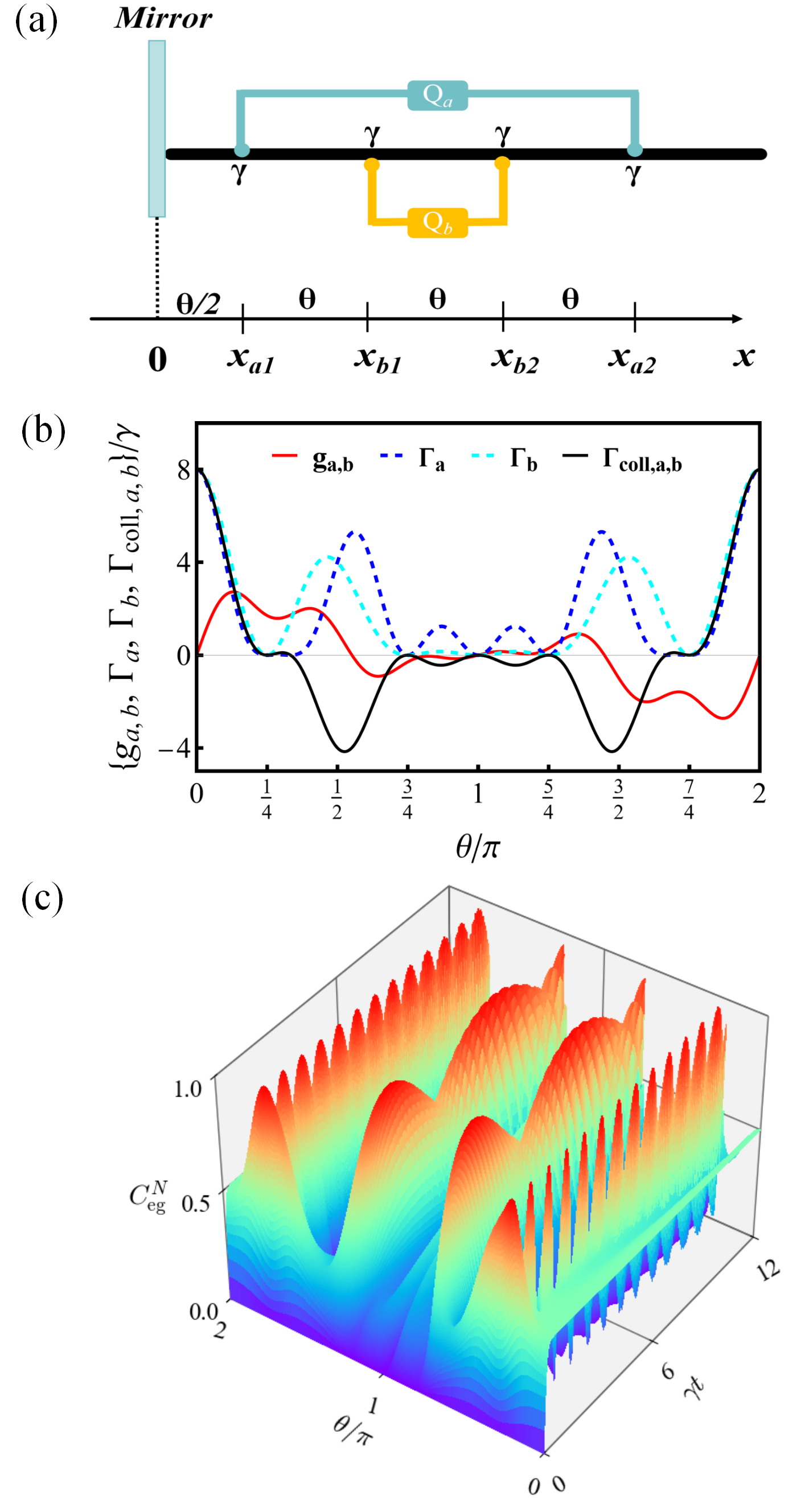}\hfill
	\caption{(Color online)  $(a)$ Two giant atoms in a nested configuration coupled to a 1D semi-infinite waveguide. $(b)$ Exchange interaction and decay rates as a function of $\theta/\pi$ for two nested giant atoms. $(c)$ Concurrence $C_{eg}^{N}$ as a function of $\theta/\pi$ and $\gamma t$. }
	\label{fig5}
\end{figure}
\section{TWO GIANT ATOMS IN SEPARATE AND NESTED CONFIGURATIONS}\label{V} 
In the above sections, we have analyzed the behavior of two braided giant atoms coupled to a semi-infinite waveguide, as well as their entanglement properties. In this section, as shown in Fig. \ref{fig4} and Fig. \ref{fig5}, we consider the separate and nested configurations. According to Eq. (\ref{eq:ex1}), the relevant coefficients in master equation (\ref{eq:me2}) for separate [Fig. \ref{fig4}$(a)$] and nested [Fig. \ref{fig5}$(a)$] configurations can be obtained. 
For the sake of conciseness, these coefficients are provided in Table \ref{tab:coefficients} in Appendix \ref{Appendix D}.

Following the strategy analogous to that used in the braided configuration, we conclude that the exchange interaction and decay rates for two giant atoms in separate and nested configurations are $2\pi$-periodic functions of $\theta$. Therefore, we restrict our discussion to the region $\theta \in [0, 2\pi]$, which fully captures the behavior of the system.
In Fig. \ref{fig4}$(b)$ and Fig. \ref{fig5}$(b)$ we show the exchange interaction and decay rates from Table \ref{tab:coefficients} in Appendix \ref{Appendix D} as a function of $\theta/\pi$, where $\theta \in [0, 2\pi]$. 

For the separate configuration, we observe from Fig. \ref{fig4}$(b)$ that there is no value of $\theta$ that satisfies the DFI condition (\ref{eq:dfcondition}). However, for the nested configuration, we note that the DFI condition (\ref{eq:dfcondition}) is satisfied when $\theta=\pi/4,  3\pi/4, 5\pi/4, 7\pi/4$, as shown in Fig. \ref{fig5}$(b)$. The result demonstrates that our scheme enables DFI in the nested configuration, challenging the prevailing notion that DFI can only be achieved in the braided configuration \cite{Kockum47, Cilluffo49, Carollo50, Soro105, Kannan48, Soro107013710, Du107023705, Raaholt6043222}. 

Similar to the braided configuration case, we generalize the study of DFI in the nested configuration to the region of $\theta \in [n\pi, (n+1)\pi]$, where $n$ is an integer. Given the periodicity of the system, the phase shift for the manifestation of DFI in the nested configuration in the region of $\theta \in [n\pi, (n+1)\pi]$ can be easily determined and is given by:
\begin{equation}
	\theta = \left(n + \frac{2k + 1}{4} \right)\pi, \quad k = 0, 1.
	\label{eq:ex12}
\end{equation}

The concurrences $C_{eg}^{S}$ and $C_{eg}^{N}$ for two giant atoms in separate and nested configurations as a function of $\theta/\pi$ and $\gamma t$ are plotted in Fig. \ref{fig4}$(c)$ and Fig. \ref{fig5}$(c)$, respectively. From Fig. \ref{fig4}$(c)$ and Fig. \ref{fig5}$(c)$, we observe that both concurrences $C_{eg}^{S}$ and $C_{eg}^{N}$ reach a steady-state value at $\theta = 0$ and $2\pi$, but drop to zero at $\theta = \pi$. These behaviors are consistent with those observed in the infinite waveguide scheme \cite{Yin55}, as presented in Appendix~\ref{Appendix C}.

In particular, the value of concurrence $C_{eg}^{S}$ can exceed 0.5 when $\theta \rightarrow \pi/2$ and $3\pi/2$, and the concurrence $C_{eg}^{N}$ is characterized by periodic oscillations ranging from zero to one at the values of $\theta$ where DFI manifests (i.e., $\theta=\pi/4, 3\pi/4, 5\pi/4, 7\pi/4$). This indicates that two nested giant atoms in the semi-infinite waveguide system can also be prepared in the maximally entangled state. Notably, in the infinite waveguide scheme \cite{Yin55}, as detailed in Appendix \ref{Appendix C}, it is not possible to generate entangled states with concurrence exceeding 0.5 in the separate configuration, nor to obtain maximally entangled states in the nested configuration. 

\section{GENERALIZATION TO MULTIPLE GIANT ATOMS}\label{VI} 
\begin{figure}[!htbp]
	\centering
	\includegraphics[width=0.45\textwidth]{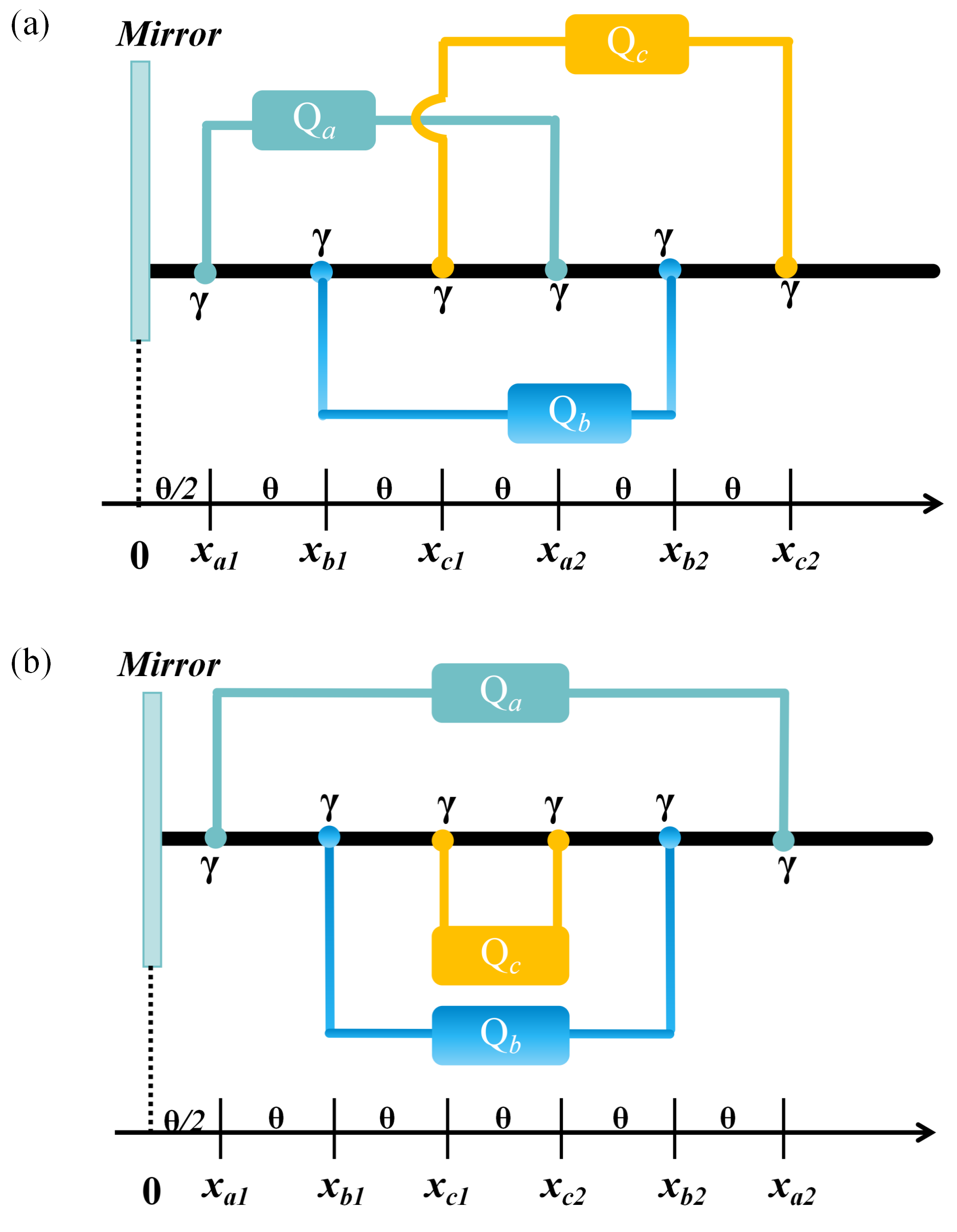}\hfill
	\caption{(Color online) Different configurations of three giant atoms: $(a)$ braided and $(b)$ nested. The notations here have the same meanings as those in Fig.~\ref{fig1}.}
	\label{fig6}
\end{figure}

Now that we have illustrated decoherence suppression and entanglement generation for two giant atoms in a semi-infinite waveguide, it is natural to generalize the discussion to multiple atoms. Here, as shown in Fig. \ref{fig6}, we consider a three-giant-atom setup, with a specific focus on the braided [Fig. \ref{fig6}$(a)$] and nested [Fig. \ref{fig6}$(b)$] configurations that have presented superior performance in Sec. \ref{III} and Sec. \ref{IV}.

\subsection{DFI}\label{VI A} 
\begin{figure}[!htbp]
	\centering
	\includegraphics[width=0.45\textwidth]{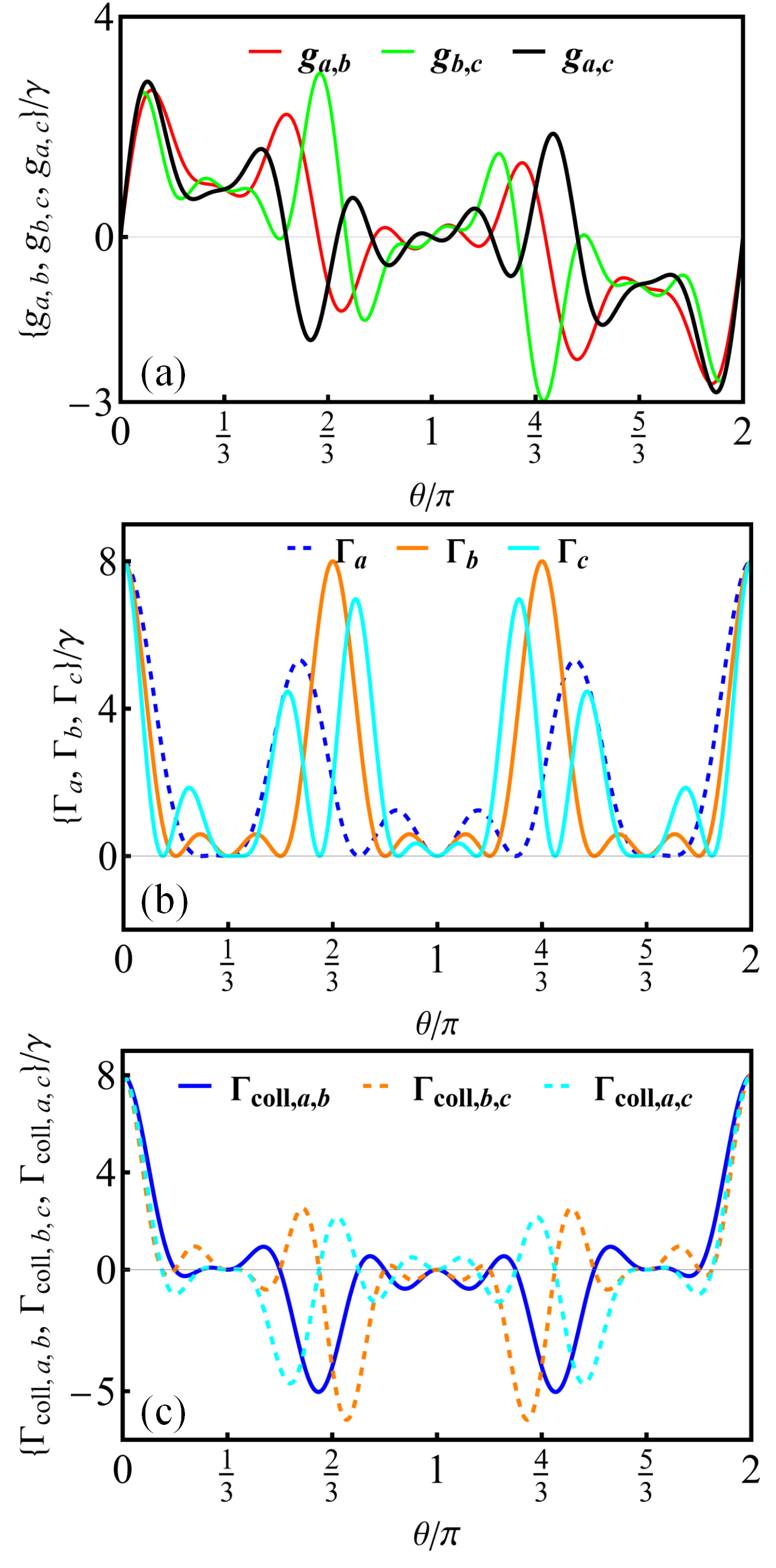}\hfill
	\caption{(Color online) $(a)$ Exchange interaction, $(b)$ individual decay rates, and $(c)$ collective decay rates as a function of $\theta/\pi$ for three braided giant atoms.}
	\label{fig7}
\end{figure}

\begin{figure}[!htbp]
	\centering
	\includegraphics[width=0.45\textwidth]{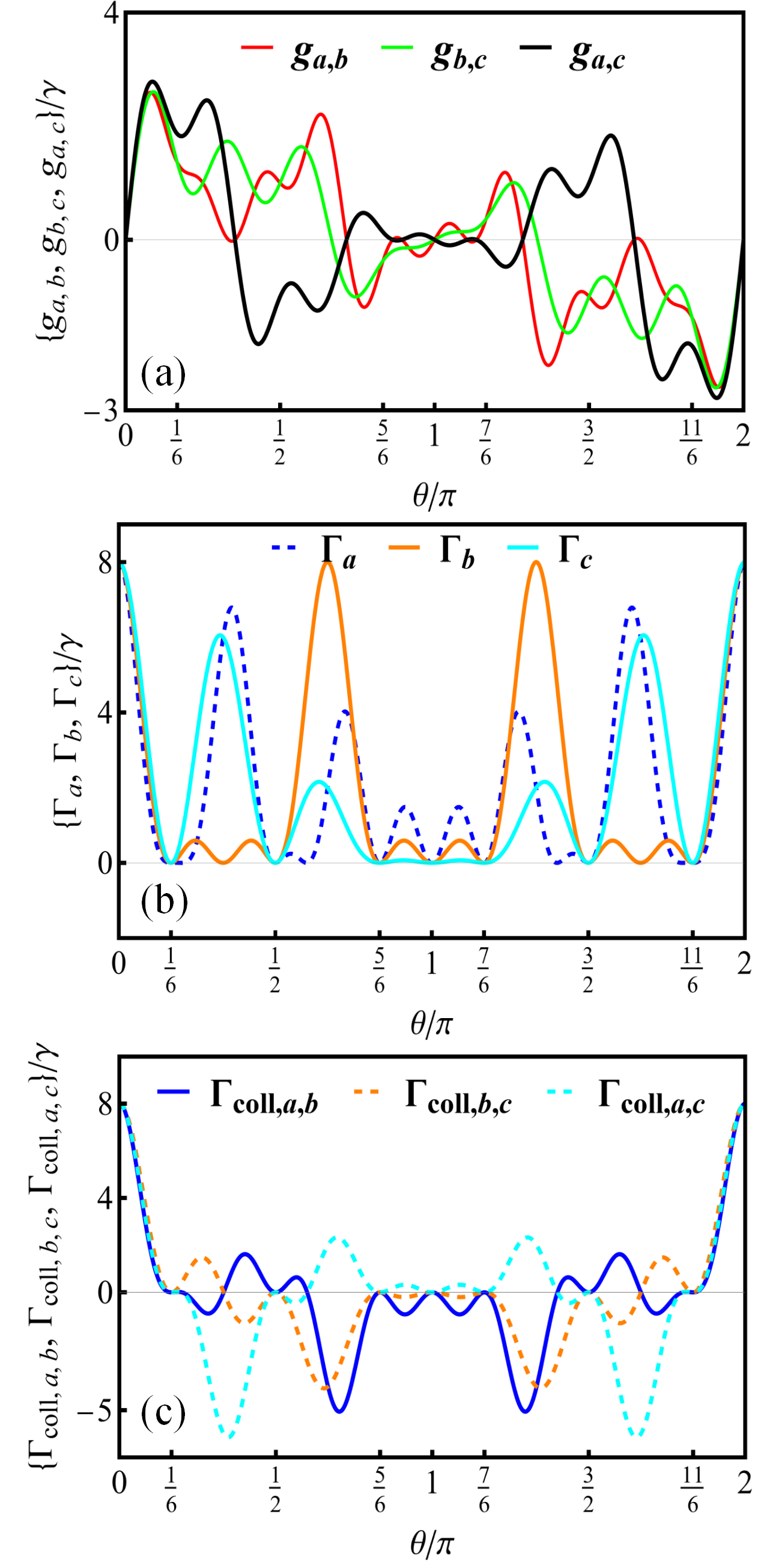}\hfill
	\caption{(Color online) $(a)$ Exchange interaction, $(b)$ individual decay rates, and $(c)$ collective decay rates as a function of $\theta/\pi$ for three nested giant atoms.}
	\label{fig8}
\end{figure}

As could be anticipated, the DFI remains feasible for multiple giant atoms in both braided and nested configurations. To demonstrate this, we follow the approach in the above section and plot the dependence of the exchange interactions and decay rates of three braided giant atoms and three nested giant atoms on $\theta$ in Figs. \ref{fig7} and \ref{fig8}, respectively. The relevant coefficient expressions can be obtained according to Eq. (\ref{eq:ex1}). By an approach analogous to the ones used in Sec. \ref{III}, we conclude that the dependence of exchange interaction and decay rates for three giant atoms in braided and nested configurations on $\theta$ is a $2\pi$-periodic function. As a result, without loss of generality, our discussion is confined to the range of $\theta \in [0,2\pi]$, which already contains all relevant features of the system. For three giant atoms, in order to achieve DFI, it is necessary to have 
\begin{equation}
	\begin{aligned}
		g_{a,b}, g_{b,c}, g_{a,c} &\neq 0, \\
		\Gamma_a, \Gamma_b, \Gamma_c &= 0, \\
		\Gamma_{coll,a,b},\Gamma_{coll,b,c}, \Gamma_{coll,a,c}&= 0.
	\end{aligned}
	\label{eq:dfcondition2}
\end{equation}

From Figs.~\ref{fig7} and \ref{fig8}, we can see that for three giant atoms in both braided and nested configurations, there are values of $\theta$ where the exchange interaction and decay rates satisfy the condition (\ref{eq:dfcondition2}), with the corresponding values given by $\theta = \pi/3, 5\pi/3$ for the braided configuration and $\theta = \pi/6, \pi/2, 5\pi/6, 7\pi/6, 3\pi/2, 11\pi/6$ for the nested configuration. These results indicate that the DFI can be realized when multiple giant atoms are in the braided and nested configurations.

Note that Refs. \cite{Kockum47, Cilluffo49, Carollo50, Soro105, Du107023705} also confirmed the manifestation of DFI between multiple giant atoms, which, however, can only be observed in the braided configuration.

\subsection{Multipartite entanglement generation}\label{VI B} 
We now turn to the entanglement generation of three giant atoms in braided and nested configurations. We let three giant atoms start in the state $|e\rangle_a |g\rangle_b|g\rangle_c$. Then the general state of the system is
\begin{equation}
	\begin{aligned}
		|\psi(t)\rangle &= c_{1}(t) |e\rangle_a |g\rangle_b |g\rangle_c + c_{2}(t) |g\rangle_a |e\rangle_b |g\rangle_c \\
		&\quad + c_{3}(t) |g\rangle_a |g\rangle_b |e\rangle_c.
	\end{aligned}
	\label{eq:Wf2}
\end{equation}

The entanglement among three giant atoms can be quantified by the linear entropy $Q$ \cite{Lambert71053804}, which is defined as
\begin{equation}
	Q \equiv \left[ \frac{1}{n} \left( \sum_{i=1}^{n} L_{i} \right) \right],
\end{equation}
where $n$ denotes the size of the system, $L_i = 2\left[1 - \text{Tr}\left(\rho_i^2\right)\right]$ represents the linear entropy of the $i$-th qubit. For the three-giant-atom quantum state given by Eq. (\ref{eq:Wf2}), the linear entropy can be expressed as \cite{Lambert71053804, Ma2024Multiparticle}
\begin{equation}
	Q = \frac{4}{3} \sum_{i=1}^{3} \left( |c_i(t)|^2 - |c_i(t)|^4 \right),
	\label{eq:entropy}
\end{equation}
where $c_1(t)$, $c_2(t)$, and $c_3(t)$ can be obtained by solving the Schrödinger equation $i \frac{\partial}{\partial t} |\psi(t)\rangle = {H}_{\text{eff}}' |\psi(t)\rangle$, with ${H}_{\text{eff}}'$ derivable via the procedure outlined in Sec. \ref{IV A}.
From Eq. (\ref{eq:entropy}), the $Q$ value of the state $W = \frac{1}{\sqrt{3}} \left( \lvert e_a g_b g_c \rangle + \lvert g_a e_b g_c \rangle + \lvert g_a g_b e_c \rangle \right)$ (a tripartite maximally entangled state) is found to be $8/9$.

In Sec. \ref{IV} and Sec. \ref{V}, we demonstrate that the two giant atoms can be prepared in a maximally entangled state when the phase shift is tuned to the value at which DFI manifests. However, for systems with multiple giant atoms, the situation is not always the same.

\begin{figure}[!htbp]
	\centering
	\includegraphics[width=0.42\textwidth]{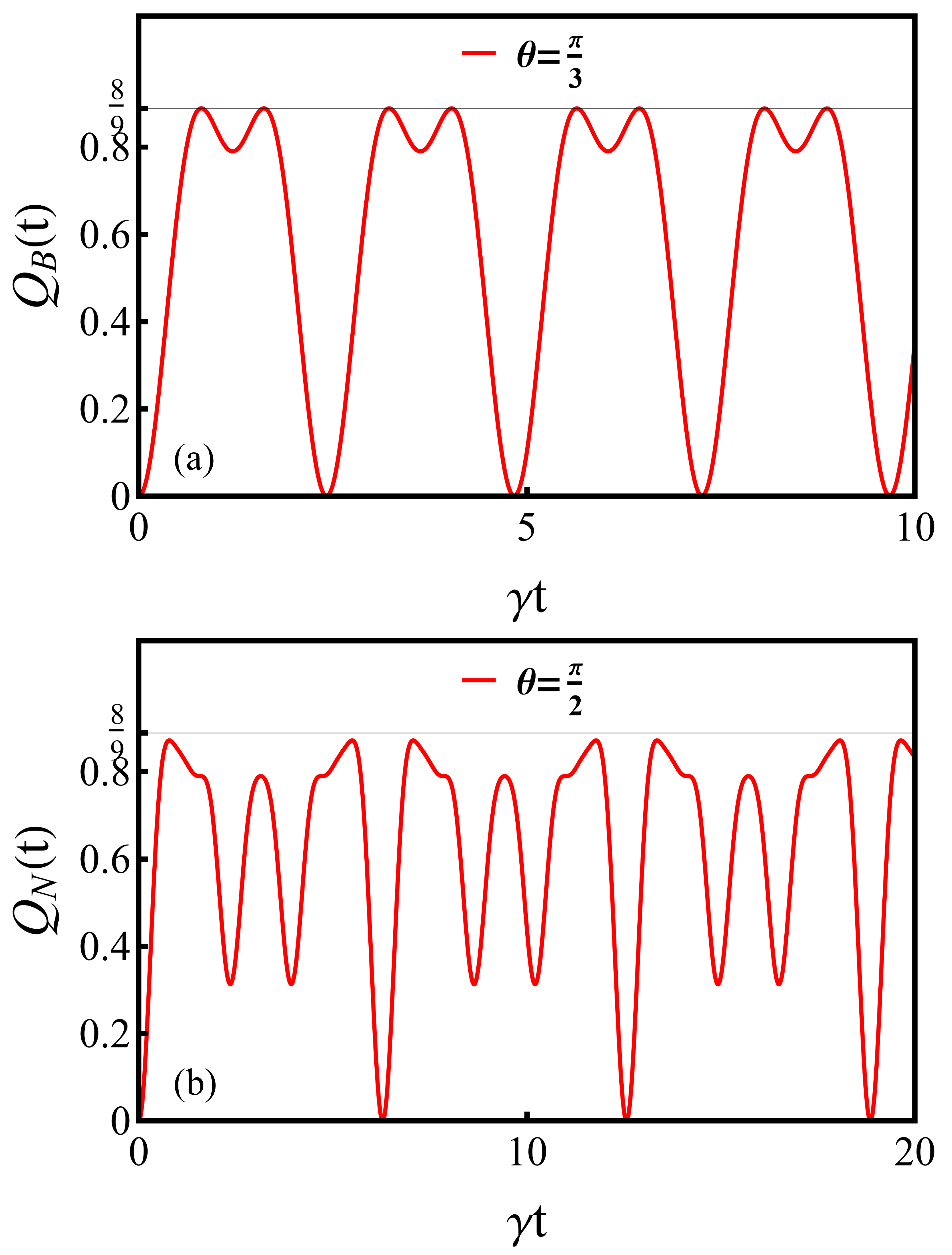}\hfill
	\caption{(Color online) Linear entropies $(a)$ $Q_{B}(t)$ and $(b)$ $Q_{N}(t)$ as a function of $\gamma t$ at a given value of phase shift.}
	\label{fig9}
\end{figure}

In Fig.~\ref{fig9}, we present the time evolution of the two linear entropies $Q_B$ and $Q_N$ at the phase-shift values where the DFI manifests, where $Q_B$ and $Q_N$ characterize the three-giant-atom entanglement in braided and nested configurations, respectively. Here, we present the results for $\theta = \pi/3$ in Fig.~\ref{fig9}$(a)$ and for $\theta = \pi/2$ in Fig.~\ref{fig9}$(b)$ as representative examples. From Fig.~\ref{fig9}$(a)$, it can be seen that at $\theta = \pi/3$, the peak value of $Q_B$ can reach 8/9, i.e., three giant atoms in the braided configuration can be prepared in the tripartite maximally entangled state--$W$ state. 
In Fig.~\ref{fig9}$(b)$, for the nested configuration, when $\theta = \pi/2$, we can see that while a high degree of entanglement can be generated among the three giant atoms, the peak value of $Q_N$ fails to reach 8/9 required for a $W$ state. That is, in the semi-infinite waveguide system with multiple giant atoms, maximally entangled states cannot always be generated even when the DFI exists between the atoms. 

\begin{figure}[!htbp]
	\includegraphics[width=0.42\textwidth]{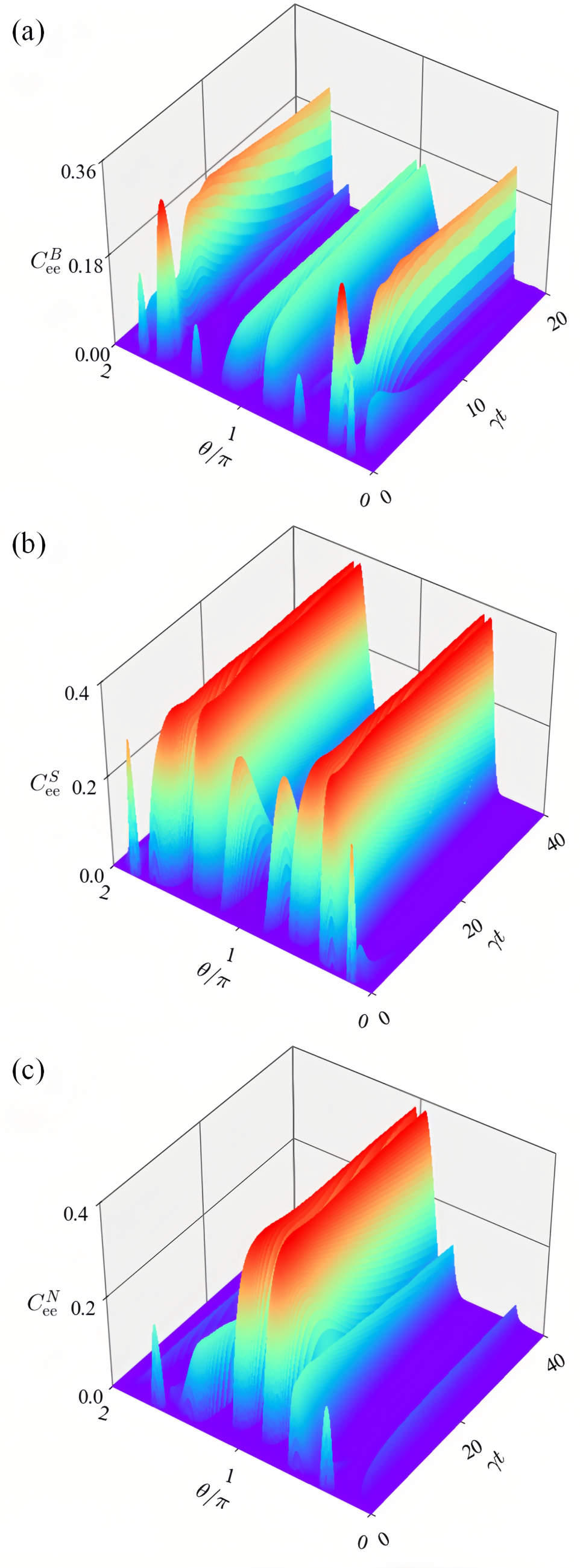}\hfill
	\caption{(Color online)  Concurrences $(a)$ $C_{ee}^{B}$, $(b)$ $C_{ee}^{S}$, and $(c)$ $C_{ee}^{N}$ as a function of $\theta/\pi$ and $\gamma t$.   }
	\label{fig10}
\end{figure}

\section{DISCUSSION}\label{VII} 
In the preceding sections, we restricted the discussion of giant-atom entanglement generation to the single-excitation subspace. We now briefly address entanglement generation for giant atoms in the two-excitation subspace. For the sake of simplicity, we focus on the case of two giant atoms. The entanglement dynamics of the two excited giant atoms across three different configurations can be obtained by numerically solving the quantum master equation (\ref{eq:me2}). As shown in Fig.~\ref{fig10}, we present the concurrences of two giant atoms with three different configurations as a function of $\theta/\pi$ and $\gamma t$ in the state $|e\rangle_a |e\rangle_b$. Figures \ref{fig10}$(a)$, \ref{fig10}$(b)$, and \ref{fig10}$(c)$ correspond to the braided, separate, and nested configurations, respectively.
The results indicate that, compared with the infinite waveguide scheme \cite{Yin55}, two excited giant atoms in the semi-infinite waveguide system can achieve stronger entanglement. To facilitate comparison, we also replot the entanglement dynamics of two excited giant atoms in three configurations reported in Ref. \cite{Yin55}, shown in Appendix \ref{Appendix C}.
From Fig.~\ref{fig10} we can see that the evolution of concurrences $C_{ee}^{B}$, $C_{ee}^{S}$, and $C_{ee}^{N}$ is also periodic with a period of \(2\pi\). Within one period, the maximum values of the concurrences $C_{ee}^{B}$, $C_{ee}^{S}$, and $C_{ee}^{N}$ are obtained to be $\approx$ 0.30, 0.38, and 0.38, respectively. In the infinite-waveguide system \cite{Yin55}, as detailed in Appendix \ref{Appendix C}, the maximum values of the concurrences for the two giant atoms in these three configurations are $\approx$ 0.029, 0.029, and 0.37, respectively. Namely, for the braided and separate configurations, the entanglement maximum value of two excited giant atoms in the semi-infinite waveguide system is enhanced by one order of magnitude compared with that in an infinite waveguide system.

We next present a discussion on the experimental feasibility of the proposed scheme. The giant atom setup has been experimentally realized recently \cite{Kannan48, Gustafsson2014, Vadiraj2021,wang2022}. In particular, Ref. \cite{Kannan48} reports a realization that two giant atoms, whose connection points are arranged in a braided manner, are coupled to an infinite microwave waveguide. Furthermore, the coupling of two small atoms to a microwave waveguide with one end terminated by a mirror has been demonstrated in experiment \cite{Wen123}. These results suggest that our scheme is achievable with current and near-future experimental conditions.

In the single-excitation initial state, we analyzed the entanglement dynamics of two giant atoms at certain phase shift values. These phase shifts are experimentally accessible by tuning the atomic frequency, as the phase shift $\theta = k_0d = \omega_0 d/v_g$ ($v_g$ represents the group velocity of the photon wave packet, $d$ is the distance between two connection points, and $\omega_0 =\omega_a=\omega_b$ is the giant atom frequency).  
For example, in Ref. \cite{Kannan48}, a phase shift of $\theta=\pi/2$ is achieved when the atomic frequency is tuned to \(4.645\,\text{GHz}\).
Using the realistic experimental parameter $\theta=\pi/2$ reported in Ref. \cite{Kannan48} for two braided giant atoms, the result shown in Fig.~\ref{fig3}$(b)$ can be realized. Though Fig.~\ref{fig3}\((b)\) is plotted up to $\gamma t=6$, the entanglement between the two braided giant atoms will keep evolving periodically between 0 and 1 when extending the time---this indicates that the entanglement can survive persistently in such periodic oscillations.

Finally, for entanglement between two atoms, it can be witnessed experimentally via violation tests of Bell's inequalities. Considering the generalized form of Bell's inequality \cite{Aspect4991, Clauser23880}:
\begin{equation}
	|S| \le 2,
\end{equation}
where 
\begin{equation}
	S = E(\vec{x},\vec{y}) - E(\vec{x},\vec{y}') + E(\vec{x}',\vec{y}) + E(\vec{x}',\vec{y}'),
\end{equation}
involves four measurements in four various orientations, and $E(\vec{x},\vec{y})$ denotes the correlation function when the two atoms are measured along orientations $\vec{x}$ and $\vec{y}$, respectively.
If experimentally measured $|S| > 2$, it demonstrates that non-local correlations exist between the two atoms, which indicates the presence of entanglement.

\section{CONCLUSION}\label{VIII} 
To summarize, we have derived the quantum master equation for multiple giant atoms coupled to a 1D semi-infinite waveguide, where each atom has multiple connection points. We specifically analyzed the behavior of two giant atoms coupled to a semi-infinite waveguide. Concretely, we considered three configurations: separate, braided, and nested. We show that two giant atoms in braided and nested configurations can interact through the semi-infinite waveguide without decohering, i.e., there is a DFI between two giant atoms. Such DFI is not possible in semi-infinite waveguide systems containing two separate giant atoms or two small atoms. This finding goes beyond the result shown in Ref. \cite{Kockum47, Cilluffo49, Carollo50, Soro105, Kannan48, Soro107013710, Du107023705} for the case of other 1D waveguides, where DFI can only be realized in the braided configuration.  

Moreover, we study the generation of entanglement between two giant atoms under three configurations in the semi-infinite waveguide system. In braided and nested configurations, the two giant atoms can reach a maximally entangled state due to the formation of DFI. In the separate configuration, the generated entanglement
between the two giant atoms can exceed 0.5.
Compared to the infinite waveguide scheme \cite{Yin55}, the maximally achievable entanglement for two giant atoms in separate and nested configurations is enhanced in the semi-infinite waveguide system.

Finally, we extend the discussion on DFI and entanglement generation to the system composed of multiple giant atoms coupled into a semi-infinite waveguide. We find that for multiple giant atoms in both braided and nested configurations, DFI can still be realized, and a high degree of entanglement between them can be generated.

This work not only verifies the possibility of realizing DFI in a semi-infinite waveguide but also provides a theoretical foundation for quantum information processing based on giant-atom semi-infinite waveguide-QED systems.

\section*{ACKNOWLEDGMENTS}
This work was supported by National Natural Science Foundation of China(Grants No. 11874190 and No. 12247101). Support was also provided by Supercomputing Center of Lanzhou University.
\begin{widetext}
\appendix

\renewcommand{\thesection}{\Alph{section}} 
\titleformat{\section}{\normalfont\normalsize\bfseries}{Appendix \thesection:}{1em}{}
\section{DERIVATION OF THE QUANTUM MASTER EQUATION FOR MULTIPLE GIANT ATOMS WITH MULTIPLE CONNECTION POINTS}\label{Appendix A}
\subsection{Two braided giant atoms with two connection points each}
\begin{figure*}[!htbp]
	\includegraphics[width=0.95\textwidth]{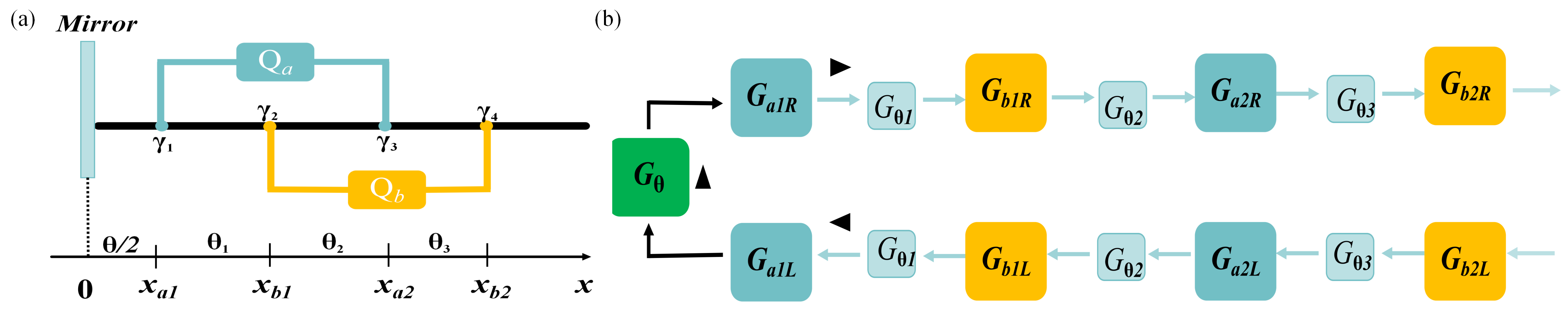}\hfill
	\caption{(Color online)  $(a)$ A sketch that shows the parameters of two braided giant atoms in a semi-infinite waveguide. $(b)$ Diagram illustrating the SLH calculation for two braided giant atoms in a semi-infinite waveguide. }
	\label{fig11}
\end{figure*}
We derived the quantum master equation for two braided giant atoms, each coupled to a 1D semi-infinite waveguide at two connection points, using the $(S, L, H)$ formalism \cite{gough58, Gough59, combes60}. In this formalism, $n$ input-output ports are needed to find the triplet for a system: S (an $n \times n$ scattering matrix), $L$ (a vector consisting of $n$ matrices that describe the system-environment interaction), and the Hamiltonian $H$ of the system (including atom-atom interaction).

Once the SLH triplet G = ($\mathbf{S}$, $\mathbf{L}$, H) for a system is calculated, the master equation for the system can be easily obtained,
\begin{align}
	\dot{\rho} = -i[H, \rho] + \sum_{k=1}^{n} \mathcal{D}[L_k]\rho,
	\label{eq:1}
\end{align}
where $\mathcal{D}[L]\rho = L\rho L^\dagger - \frac{1}{2} \left( L^\dagger L \rho + \rho L^\dagger L \right)$ are Lindblad operator.

To achieve this, we present the two braided ginat atom setup and corresponding schematic representation of the SLH calculation in Fig. \ref{fig11}.
The triplets for the right moving part and the left moving part are respectively

\begin{align}
	G_{\text{R}} &= G_{b\text{2R}} \triangleleft G_{\theta_3} \triangleleft G_{a\text{2R}} \triangleleft G_{\theta_2} \triangleleft G_{b\text{1R}} \triangleleft G_{\theta_1} \triangleleft G_{a\text{1R}}, \notag \\
	G_{\text{L}} &= G_{a\text{1L}} \triangleleft G_{\theta_1} \triangleleft G_{b\text{1L}} \triangleleft G_{\theta_2} \triangleleft G_{a\text{2L}} \triangleleft G_{\theta_3} \triangleleft G_{b\text{2L}},
	\label{eq:2}
\end{align}
where
\begin{equation}
	\begin{alignedat}{3}
		G_{a1R} &= \left( 1, \sqrt{\frac{\gamma_1}{2}} \sigma_{-}^a, \omega_a \frac{\sigma_z^a}{2} \right), 
		&\quad G_{a2R} &= \left( 1, \sqrt{\frac{\gamma_3}{2}} \sigma_{-}^a, 0 \right), \\
		G_{a1L} &= \left( 1, \sqrt{\frac{\gamma_1}{2}} \sigma_{-}^a, 0 \right), 
		&\quad G_{a2L} &= \left( 1, \sqrt{\frac{\gamma_3}{2}} \sigma_{-}^a, 0 \right), \\
		G_{b1R} &= \left( 1, \sqrt{\frac{\gamma_2}{2}} \sigma_{-}^b, \omega_b \frac{\sigma_z^b}{2} \right), 
		&\quad G_{b2R} &= \left( 1, \sqrt{\frac{\gamma_4}{2}} \sigma_{-}^b, 0 \right), \\
		G_{b1L} &= \left( 1, \sqrt{\frac{\gamma_2}{2}} \sigma_{-}^b, 0 \right), 
		&\quad G_{b2L} &= \left( 1, \sqrt{\frac{\gamma_4}{2}} \sigma_{-}^b, 0 \right). 
	\end{alignedat}
	\label{eq:3}
\end{equation}
and the phase shift acquired is represented by $G_{\theta_k} = (e^{i\theta_k}, 0, 0)$, where $k = 1, 2, 3$ (Specifically, $G_{\theta} = (e^{i\theta}, 0, 0)$). 
The connection between two SLH triplets, symbolized as \( G_2 \triangleleft G_1 \), results in the following outcome \cite{gough58, Gough59, combes60}:
\begin{equation}
	\begin{aligned} 
		G_2 \triangleleft G_1 
		&= (\mathbf{S_2}, \mathbf{L_2}, H_2) \triangleleft (\mathbf{S_1}, \mathbf{L_1}, H_1) \\ 
		&= \left( \mathbf{S}_2 \mathbf{S}_1, \mathbf{S}_2 \mathbf{L}_1 + \mathbf{L}_2, H_1 + H_2 + \frac{1}{2i} \left[ \mathbf{L}_2^\dagger \mathbf{S}_2 \mathbf{L}_1 - \mathbf{L}_1^\dagger \mathbf{S}_2^\dagger \mathbf{L}_2 \right] \right).
	\end{aligned}
	\label{eq:4}
\end{equation}
Additionally, each SLH triplet in Eq. (\ref{eq:3}) is given according to the definition of SLH triplet at each connection point \cite{Kockum46}: 
\begin{align}  
	G_{R} &= \begin{cases} 
		(1, \sqrt{\frac{\gamma_{jn}}{2}} \sigma_{-}^j , \frac{1}{2} \omega_j \sigma_z^j) & \quad \text{if } n \text{ is the first connection point of atom } j, \\
		(1, \sqrt{\frac{\gamma_{jn}}{2}} \sigma_{-}^j, 0) & \quad \text{otherwise}.
	\end{cases} \notag \\ 
	G_{L} &= \begin{cases} 
		(1, \sqrt{\frac{\gamma_{jn}}{2}} \sigma_{-}^j , 0) & \hspace{1.2cm} \text{applicable to all connection points of atom } j.
	\end{cases}  
	\tag{A.5} 
\end{align}

Then, the total triplet for the system is
\begin{align}
	G_{\text{tot}} &= ( \mathbf{S_{tot}}, \mathbf{L_{tot}}, H_{tot} ) = G_{\text{R}} \triangleleft G_{\theta} \triangleleft G_{\text{L}}, 
	\tag{A.6} 
\end{align}
with
\begin{equation}
	\begin{aligned}
		S_{\text{tot}} &= e^{i(2\theta_1 + 2\theta_2 + 2\theta_3 + \theta)}, \\
		L_{\text{tot}} &= 
		\left ((e^{i(\theta_1 + \theta_2 + \theta_3 + \theta )}  + e^{i(\theta_1 + \theta_2 + \theta_3 )} )\sqrt{\frac{\gamma_1}{2}} + (e^{i(2\theta_1 + 2\theta_2 + \theta_3 + \theta)}  + e^{i \theta_3}) \sqrt{\frac{\gamma_3}{2}} \right ) \sigma_{-}^{a} \\
		&+ \left ((e^{i(2\theta_1 + \theta_2 + \theta_3 + \theta )}  + e^{i( \theta_2 + \theta_3 )}) \sqrt{\frac{\gamma_2}{2}} + (e^{i(2\theta_1 + 2\theta_2 + 2\theta_3 + \theta )}  +1 )\sqrt{\frac{\gamma_4}{2}} \right)\sigma_{-}^{b} , \\
		H_{\text{tot}} &= \omega_a \frac{\sigma_z^a}{2} +  \omega_b \frac{\sigma_z^b}{2} + \left[ \sqrt{\gamma_1 \gamma_3} (\sin(\theta_1 + \theta_2) + \sin(\theta_1 + \theta_2 + \theta)) +\frac{\gamma_1}{2} \sin \theta + \frac{\gamma_3}{2} \sin(2\theta_1 + 2\theta_2 + \theta) \right] \sigma_{+}^{a} \sigma_{-}^{a} \\
		&+ \left[ \sqrt{\gamma_2 \gamma_4} (\sin(\theta_2 + \theta_3) + \sin(2\theta_1 + \theta_2 + \theta_3 + \theta)) +\frac{\gamma_2}{2} \sin( 2\theta_1 + \theta ) + \frac{\gamma_4}{2} \sin(2\theta_1 + 2\theta_2 + 2\theta_3 + \theta) \right] \sigma_{+}^{b} \sigma_{-}^{b} \\
		&+ \left[ \frac{\sqrt{\gamma_1 \gamma_2}}{2} ( \sin \theta_1 + \sin(\theta_1 + \theta) ) 
		+\frac{\sqrt{\gamma_2 \gamma_3}}{2} ( \sin \theta_2 + \sin(2\theta_1 +\theta_2 + \theta) ) \right. \notag \\
		&\left. +\frac{\sqrt{\gamma_3 \gamma_4}}{2} ( \sin \theta_3 + \sin(2\theta_1 +2\theta_2 + \theta_3 + \theta)) 
		+\frac{\sqrt{\gamma_1 \gamma_4}}{2} ( \sin (\theta_1 +\theta_2 + \theta_3 ) + \sin(\theta_1 +\theta_2 + \theta_3+ \theta) ) \right] \notag \\
		&\quad \left( \sigma_{+}^{b} \sigma_{-}^{a} + \sigma_{+}^{a} \sigma_{-}^{b} \right). \\
	\end{aligned}
	\tag{A.7}
	\label{eq:7}
\end{equation}

Using the following properties of the $\bm{L}$ operator:
\begin{align}
	D[a + b] \rho &=  (a + b)\rho(a^\dagger + b^\dagger) - \frac{1}{2} \left[ (a^\dagger + b^\dagger)(a + b)\rho + \rho(a^\dagger + b^\dagger)(a + b) \right] \notag \\
	&= a\rho a^\dagger + a\rho b^\dagger + b\rho a^\dagger + b\rho b^\dagger - \frac{1}{2} \left[ (a^\dagger a + a^\dagger b + b^\dagger a + b^\dagger b)\rho + \rho(a^\dagger a + a^\dagger b + b^\dagger a + b^\dagger b) \right] \notag \\
	&=D[a] \rho + D[b] \rho + a \rho b^\dagger + b \rho a^\dagger - \frac{1}{2} \left(a^\dagger b + b^\dagger a\right) \rho - \frac{1}{2} \rho \left(a^\dagger b + b^\dagger a\right).
	\tag{A.8}
	\label{eq:8}
\end{align}

The master equation finally reads as
\begin{align}
	\dot{\rho} &= -i[H_{\text{tot}}, \rho] + \mathcal{D} \left[L_{\text{tot}} \right] \rho \notag \\ 
	&= -i[H_{\text{tot}}, \rho] + \mathcal{D} \left[\left( (e^{i(\theta_1 + \theta_2 + \theta_3 + \theta)}  + e^{i(\theta_1 + \theta_2 + \theta_3)}) \sqrt{\frac{\gamma_1}{2}} 
	+ (e^{i(2\theta_1 + 2\theta_2 + \theta_3 + \theta)}  + e^{i \theta_3}) \sqrt{\frac{\gamma_3}{2}} \right) \sigma_{-}^{a} \right. \notag \\
	&\quad + \left. \left( (e^{i(2\theta_1 + \theta_2 + \theta_3 + \theta)}  + e^{i( \theta_2 + \theta_3)}) \sqrt{\frac{\gamma_2}{2}} 
	+ (e^{i(2\theta_1 + 2\theta_2 + 2\theta_3 + \theta)}  + 1 ) \sqrt{\frac{\gamma_4}{2}} \right) \sigma_{-}^{b} \right] \rho \notag \\ 
	&= -i[H_{\text{tot}}, \rho] +\left[ \gamma_{1} + \gamma_{3} + \gamma_{1} \cos\theta + \gamma_{3} \cos(2\theta_1 + 2\theta_2 + \theta) + 2 \sqrt{\gamma_{1} \gamma_{3}} \cos(\theta_1 + \theta_2) + 2 \sqrt{\gamma_{1} \gamma_{3}} \cos(\theta_1 + \theta_2 + \theta ) \right] \mathcal{D}[\sigma_-^{a}] \rho \notag \\
	&+ \left[ \gamma_{2} + \gamma_{4} + \gamma_{2} \cos(2\theta_1 + \theta) + \gamma_{4} \cos(2\theta_1 + 2\theta_2 + 2\theta_3 + \theta) + 2 \sqrt{\gamma_{2} \gamma_{4}} \cos(\theta_2 + \theta_3) + 2 \sqrt{\gamma_{2} \gamma_{4}} \cos(2\theta_1 + \theta_2 + \theta_3 + \theta ) \right] \mathcal{D}[\sigma_-^{b}] \rho \notag \\
	&+ \left[ \sqrt{\gamma_{1} \gamma_{2}} (\cos \theta_1 + \cos(\theta_1 + \theta)) 
	+ \sqrt{\gamma_{2} \gamma_{3}} (\cos \theta_2 + \cos(2\theta_1 + \theta_2 + \theta)) \right. \notag \\
	& \left. + \sqrt{\gamma_{3} \gamma_{4}} (\cos \theta_3 + \cos(2\theta_1 + 2\theta_2 + \theta_3 + \theta)) 
	+ \sqrt{\gamma_{1} \gamma_{4}} (\cos (\theta_1 + \theta_2 + \theta_3) + \cos(\theta_1 + \theta_2 + \theta_3 + \theta)) \right] \notag \\
	& \quad \times \left[ (\sigma_{-}^{a}\rho\sigma_{+}^{b} - \frac{1}{2} \left\{ \sigma_{+}^{b}\sigma_{-}^{a}, \rho \right\}) + \text{H.c.}\right].
	 \tag{A.9} 
	\label{eq:9}
\end{align}

In particular, when Eq. (\ref{eq:9}) is transformed into the interaction picture and the assumptions $\gamma_{1} = \gamma_{2} = \gamma_{3} = \gamma_{4}= \gamma$, $\theta_1 = \theta_2 = \theta_3 = \theta$ are made, it reduces to Eq. (\ref{eq:me2}) of the main text, with the coefficients given in Eq. (\ref{eq:ex2}).
\subsection{Multiple giant atoms with multiple connection points each}
The master equation for $M$ giant atoms coupled to a semi-infinite waveguide, where each atom $j$ has $N_j$ connection points, can be derived by following the same procedure described above. The total SLH triplet for the system is given by
\begin{equation}
	\begin{aligned}
		S_{\text{tot}} &= e^ {ik_0\left[|x_{1_{1}}-x_{M_{N}}|+ (x_{1_{1}}+x_{M_{N}}) \right]}, \\
		L_{\text{tot}} &= 
		\sum_{j=1}^{M} \sum_{n=1}^{N_j} \Big\{ e^ {ik_0|x_{jn}-x_{M_{N}}|} + e^ {ik_0(x_{jn}+x_{M_{N}})}\Big\} \sqrt{\frac{\gamma_{jn}}{2}} \sigma_{-}^{j}, \\
		H_{\text{tot}} &= \sum_{j=1}^{M} \omega_j \frac{\sigma_z^j}{2}+ \sum_{j=1}^{M} \sum_{n=1}^{N_j} \sum_{m=1}^{N_j} \frac{\sqrt{\gamma_{jn} \gamma_{jm}}}{2} 
		\Big\{ \sin \left[k_0|x_{jn}-x_{jm}|\right] +  \sin \left[k_0(x_{jn}+x_{jm})\right] \Big\} \sigma_{+}^{j} \sigma_{-}^{j}\\
		&+\sum_{j=1}^{M-1} \sum_{k=j+1}^{M} \sum_{n=1}^{N_j} \sum_{m=1}^{N_k} \frac{\sqrt{\gamma_{jn} \gamma_{km}}}{2} 
		\Big\{\sin \left[k_0|x_{jn}-x_{km}|\right] + \sin \left[k_0(x_{jn}+x_{km})\right] \Big\}
		\left(  \sigma_+^{(k)}  \sigma_-^{(j)} +\text{H.c.} \right), \\
	\end{aligned}
	\tag{A10}
	\label{eq:10}
\end{equation}
where $x_{1_{1}}$ and $x_{M_{N}}$ denote the coordinates of the connection points that are closest to and farthest from the mirror respectively.

From the SLH triplet above, the Lindblad master equation including $M$ giant atoms can be formulated as
\begin{align}
	\dot{\rho} &= -i[H_{\text{tot}}, \rho] + \mathcal{D} \left[ L_{tot} \right] \rho \notag \\
	&= -i[H_{\text{tot}}, \rho] + \mathcal{D} \left[\sum_{j=1}^{M} \sum_{n=1}^{N_j} \Big\{ e^ {ik_0|x_{jn}-x_{M_{N}}|} + e^ {ik_0(x_{jn}+x_{M_{N}})}\Big\} \sqrt{\frac{\gamma_{jn}}{2}} \sigma_{-}^{j} \right] \rho .
	\tag{A11}
	\label{eq:11}
\end{align}
By expanding the second term on the right-hand side of Eq. (\ref{eq:11}) with the properties of the $\bm{L}$ operator
\begin{equation}
	D\left[ \sum_{j=1}^{M} O_j \right] \rho = \sum_{j=1}^{M} \sum_{i=1}^{M} O_j \rho O_i^\dagger - \frac{1}{2} \sum_{j=1}^{M} \sum_{i=1}^{M} \left( O_i^\dagger O_j \rho + \rho O_i^\dagger O_j \right),
	\tag{A12}
	\label{eq:12}
\end{equation}
we finally arrive at
\begin{align}
	\dot{\rho} &= -i[H_{\text{tot}}, \rho] + \mathcal{D} \left[\sum_{j=1}^{M} \sum_{n=1}^{N_j} \Big\{ e^ {ik_0|x_{jn}-x_{M_{N}}|} + e^ {ik_0(x_{jn}+x_{M_{N}})}\Big\} \sqrt{\frac{\gamma_{jn}}{2}} \sigma_{-}^{j} \right] \rho \notag \\
	&= -i[H_{\text{tot}}, \rho] + \sum_{j=1}^{M}  \sum_{n=1}^{N_j} \sum_{m=1}^{N_j} \sqrt{\gamma_{j_{n}} \gamma_{j_{m}}} \Big\{ \cos \left[k_0|x_{jn}-x_{jm}|\right] +  \cos \left[k_0(x_{jn}+x_{jm})\right] \Big\} \mathcal{D}[\sigma_-^{j}] \rho \notag \\
	&+\sum_{j=1}^{M-1} \sum_{k=j+1}^{M} \sum_{n=1}^{N_j} \sum_{m=1}^{N_k} \sqrt{\gamma_{j_{n}} \gamma_{k_{m}}} \Big\{\cos \left[k_0|x_{jn}-x_{km}|\right] + \cos \left[k_0(x_{jn}+x_{km})\right] \Big\}
	\left[ \left( \sigma_-^{(j)} \rho \sigma_+^{(k)} - \frac{1}{2} \left\{ \sigma_+^{(k)} \sigma_-^{(j)}, \rho \right\} \right)+\text{H.c.} \right].
	\tag{A13}
	\label{eq:13}
\end{align}
When we transform Eq. (\ref{eq:13}) into the interaction picture, it reduces to Eq. (\ref{eq:me1}) in the main text, with the corresponding coefficients provided in Eq. (\ref{eq:ex1}).

\section{ANALYTICAL FORMS OF THE PHASE-DEPENDENT FUNCTIONS $R_i(t)$ $(i=1,2,3,4)$ IN EQ.~(\ref{eq:Concurrence})}\label{Appendix B}
Expressions of the functions $R_i(t)$ $(i=1,2,3,4)$ in Eq.~(\ref{eq:Concurrence}) have the following form:
	\begin{equation}
		\begin{aligned}  
			R_1(t) &=\gamma t\left(D + 2A \cos(\frac{\theta }{2})\right),\\
			R_2(t) &= E \left(  e^{\sqrt{2}\, B^* \gamma t \cos\left( \frac{\theta}{2} \right)} -1 \right) , \\
			R_3(t) &= 2 A \left(  e^{\sqrt{2}\, B \gamma t \cos\left( \frac{\theta}{2} \right)} +1 \right) , \\
			R_4(t) &= F \left( e^{\sqrt{2}\, B \gamma t \cos\left( \frac{\theta}{2} \right)} - 1 \right) ,
		\end{aligned}
	\end{equation}
	with
	\begin{equation}
		\begin{aligned} 
			A &= \sqrt{ e^{3i\theta} \left( 37 - 50 e^{i\theta} + 93 e^{2i\theta} - 80 e^{3i\theta} + 86 e^{4i\theta} 
				- 52 e^{5i\theta} + 38 e^{6i\theta} - 16 e^{7i\theta} + 9 e^{8i\theta} - 2 e^{9i\theta} + e^{10i\theta} \right) }, \\
			D &= 4 + e^{i\theta} + 4 e^{2i\theta} + 3 e^{3i\theta} + 3 e^{5i\theta} + e^{7i\theta}, \\ 
			E &=  -3 + \cos{\theta}  \left( 3 - 4i \sin{\theta} \right) + i \sin{\theta} - 4 \cos{2\theta} , \\
			F &= 16 i e^{4i\theta} \cos^2{\theta} \sin\left( \frac{\theta}{2} \right),\\
			B & =\sqrt{(B_1+B_2)e^{8i\theta}} .
		\end{aligned}
	\end{equation}
	Here, $B_1$ and $B_2$ are defined as
	\begin{equation}
		\begin{aligned} 
	B_1 &=  -26 + 62 \cos{\theta} - 48 \cos{2\theta} + 51 \cos{3\theta} - 26 \cos{4\theta} + 19 \cos{5\theta},  \\		
	B_2 &= -24i \sin{\theta} + 32i \sin{2\theta} - 42i \sin{3\theta} + 24i \sin{4\theta} - 18i \sin{5\theta}.  \\
\end{aligned}
\end{equation}
\section{THE ENTANGLEMENT GENERATION BETWEEN TWO GIANT ATOMS WITH THREE DIFFERENT CONFIGURATIONS IN INFINITE WAVEGUIDE SYSTEMS}\label{Appendix C}
In this appendix, we replot the corresponding results for the infinite waveguide case, as shown in Fig.~\ref{fig12}.
\begin{figure*}[!htbp]
	\centering
	\includegraphics[width=0.90\textwidth]{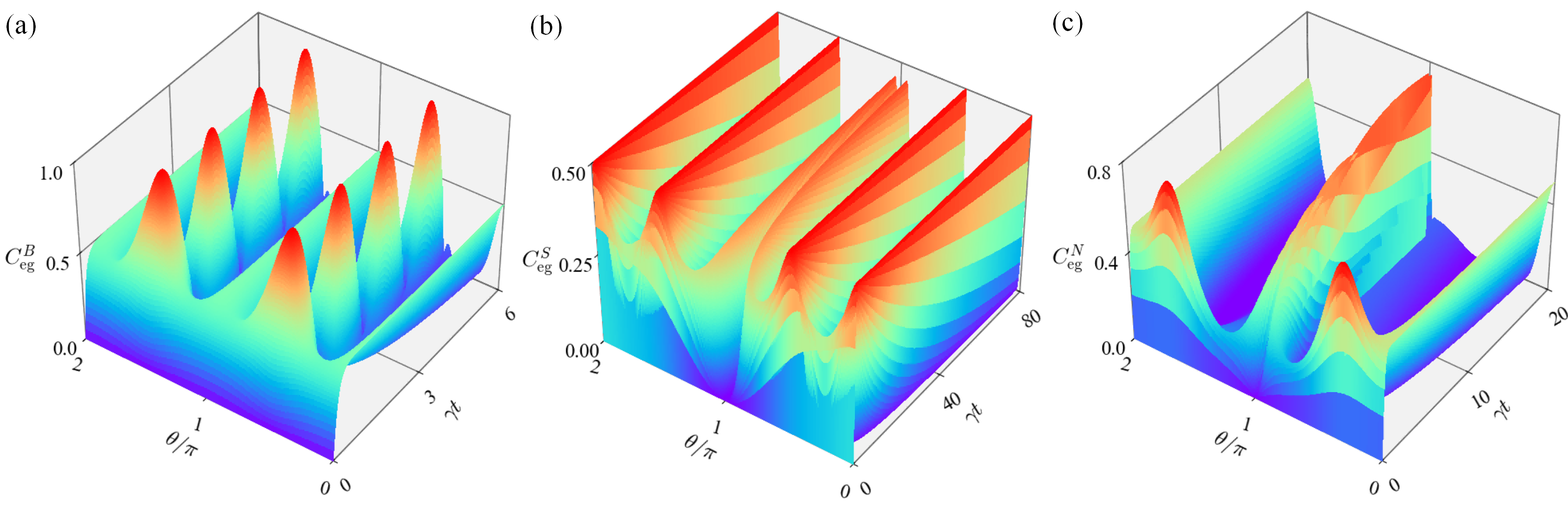}\hfill
	\includegraphics[width=0.90\textwidth]{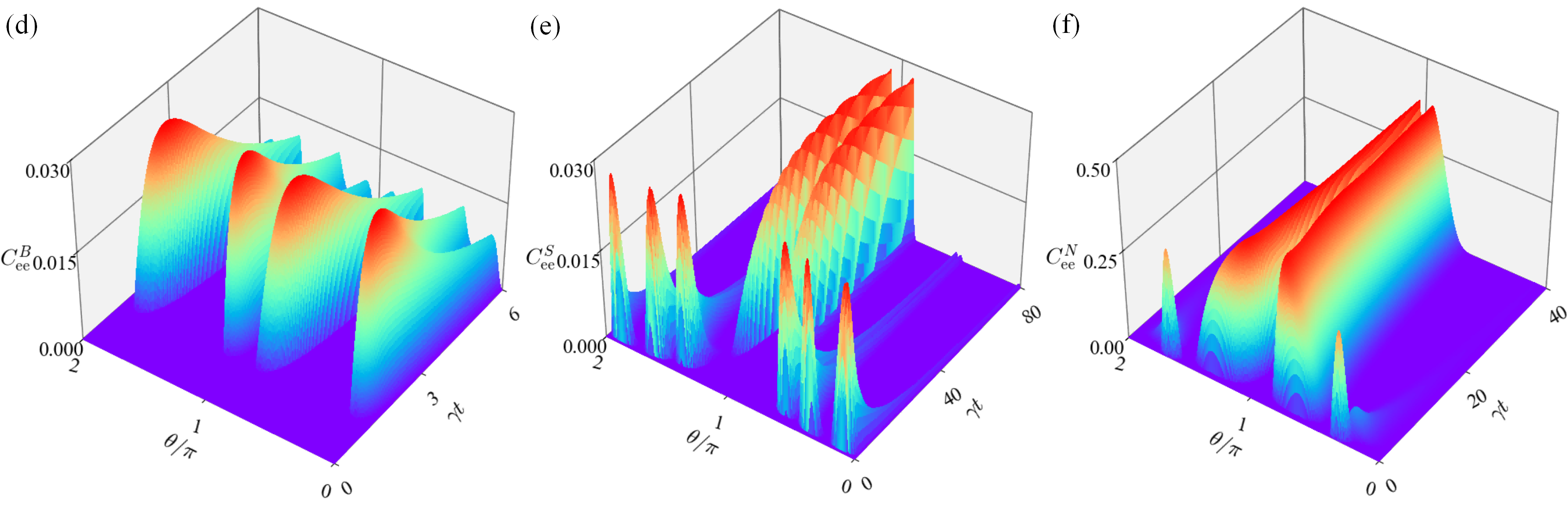}\hfill
	\caption{(Color online) Concurrences as a function of $\theta/\pi$ and $\gamma t$. In the upper and lower columns, atomic initial state $|\psi(0)\rangle = |e\rangle_a |g\rangle_b$ and $ |e\rangle_a |e\rangle_b$, respectively. The two giant atoms are arranged in braided [$(a)$, $(d)$], separate [$(b)$, $(e)$], and nested configurations [$(c)$, $(f)$]. }
	\label{fig12}
\end{figure*}

\section{COEFFICIENT EXPRESSIONS FOR TWO GIANT ATOMS IN SEPARATE AND NESTED CONFIGURATIONS}\label{Appendix D}

\begin{table}[H]
	\caption{Lamb shifts ($\delta\omega_a$, $\delta\omega_b$), exchange interaction ($g_{a,b}$), individual decay rates ($\Gamma_a$, $\Gamma_b$), and collective decay rates ($\Gamma_{coll,a,b}$) for two giant atoms in separate and nested configurations.}
	\label{tab:coefficients}
	\small
	\setlength{\tabcolsep}{6pt} 
	
	\begin{tabularx}{\linewidth}{llXX}
		\hline\hline
		Coefficient & Configuration & Atom $a$ & Atom $b$ \\
		\hline
		$\delta\omega_a,\delta\omega_b$
		& Separate & $\gamma \sin \theta + \gamma \sin (2\theta) + \frac{\gamma}{2} \sin \theta + \frac{\gamma}{2} \sin (3\theta)$
		& $\gamma \sin \theta + \gamma \sin (6\theta) + \frac{\gamma}{2} \sin (5\theta) + \frac{\gamma}{2} \sin (7\theta)$ \\
		& Nested   & $\gamma \sin (3\theta) + \gamma \sin (4\theta) + \frac{\gamma}{2} \sin \theta + \frac{\gamma}{2} \sin (7\theta)$
		& $\gamma \sin \theta + \gamma \sin (4\theta) + \frac{\gamma}{2} \sin (3\theta) + \frac{\gamma}{2} \sin (5\theta)$ \\
		\hline
		$\Gamma_a,\Gamma_b$
		& Separate 
		& $2\gamma + 2\gamma \cos\theta + 2\gamma \cos(2\theta) + \gamma \cos \theta$ 
		& $2\gamma + 2\gamma \cos\theta + 2\gamma \cos(6\theta)+\gamma\cos(5\theta)$ \\
		& 
		& $+\gamma \cos(3\theta)$
		& $+ \gamma \cos(7\theta)$ \\
		& Nested   
		& $2\gamma + 2\gamma \cos(3\theta) +2\gamma \cos(4\theta)+ \gamma \cos\theta $ 
		& $2\gamma + 2\gamma \cos\theta + 2\gamma \cos(4\theta)+\gamma\cos(3\theta) $ \\
		& 
		& $\gamma \cos(7\theta)$
		& $+ \gamma \cos(5\theta)$ \\
		\hline
		$\Gamma_{coll,a,b}$ 
		& Separate & $\gamma\!\left[\cos\theta+2\cos(2\theta)+2\cos(3\theta)+2\cos(4\theta)+\cos(5\theta)\right]$ & \\
		& Nested  & $\gamma\!\left[2\cos\theta+3\cos(2\theta)+\cos(3\theta)+\cos(5\theta)+\cos(6\theta)\right]$ & \\
		\hline
		$g_{a,b}$
		& Separate
		& $\frac{\gamma}{2}\!\left[
		\sin\theta+2\sin(2\theta)+2\sin(3\theta)
		+2\sin(4\theta)+\sin(5\theta)
		\right]$
		& \\
		& Nested
		& $\frac{\gamma}{2}\!\left[
		2\sin\theta+3\sin(2\theta)+\sin(3\theta)+\sin(5\theta)+\sin(6\theta)
		\right]$
		& \\
		\hline\hline
	\end{tabularx}
\end{table}

\end{widetext}
\bibliography{REF}

\end{document}